\documentclass[aps,prb,superscriptaddress,amsmath,amssymb,
floatfix,twocolumn]{revtex4-2}

\usepackage{graphicx}
\usepackage{physics}
\usepackage{braket}
\usepackage{times}
\usepackage{multirow}
\usepackage{ascmac}
\usepackage{amsthm}
\usepackage{float}
\usepackage{color}
\usepackage{comment}
\usepackage{mathtools}  
\usepackage{mathrsfs} 
\usepackage[colorlinks=True,urlcolor=blue,linkcolor=blue,citecolor=blue]{hyperref}
\usepackage{xcolor}
\usepackage{multirow}
\usepackage{setspace}
\usepackage{bm}

\usepackage[toc,page]{appendix}

\usepackage{amsfonts}
\usepackage{amsmath}
\usepackage{amssymb}
\usepackage{bm}

\usepackage{slashed}
\usepackage{dsfont}

\begin{document}
	
	\title{Interplay between non-Fermi liquid and non-Hermiticity: A multi-method study of non-Hermitian multichannel Kondo model}
	
	\author{Wei-Zhu Yi}
	\affiliation{National Laboratory of Solid State Microstructures and Department of Physics, Nanjing University, Nanjing 210093, China}
	\affiliation{State Key Laboratory of Quantum Functional Materials and Department of Physics, Southern University of Science and Technology, Shenzhen 518055,  China}
	\author{Yun Chen}
	\affiliation{National Laboratory of Solid State Microstructures and Department of Physics, Nanjing University, Nanjing 210093, China}
	\author{Jun-Jun Pang}
	\affiliation{National Laboratory of Solid State Microstructures and Department of Physics, Nanjing University, Nanjing 210093, China}
	\author{Hong Chen}
	\affiliation{National Laboratory of Solid State Microstructures and Department of Physics, Nanjing University, Nanjing 210093, China}
	\author{Baigeng Wang}
	\affiliation{National Laboratory of Solid State Microstructures and Department of Physics, Nanjing University, Nanjing 210093, China}
	\affiliation{Collaborative Innovation Center of Advanced Microstructures, Nanjing University, Nanjing 210093, China}
	\affiliation{Jiangsu Physical Science Research Center, Nanjing University, Nanjing 210093, China}
	\author{Rui Wang}
	\email{rwang89@nju.edu.cn}
	\affiliation{National Laboratory of Solid State Microstructures and Department of Physics, Nanjing University, Nanjing 210093, China}
	\affiliation{Collaborative Innovation Center of Advanced Microstructures, Nanjing University, Nanjing 210093, China}
	\affiliation{Jiangsu Physical Science Research Center, Nanjing University, Nanjing 210093, China}
	\affiliation{Hefei National Laboratory, Hefei 230088, People's Republic of China }

	\date{\today}
	
	\begin{abstract}
		Non-Hermitian multichannel Kondo problems host both non-Fermi liquid and non-Hermitian physics, which provide a prototypical model to explore exotic collective quantum phenomena driven by the two different ingredients. Here, we first propose an experimental setup that realizes this model with exact channel symmetry as well as a controllable \(\mathcal{PT}\) symmetry. Then, we perform a multi-method study of this model, focusing on the low-energy spectrum, the thermodynamic quantities, and the transport properties associated with different fixed points. Using the Bethe ansatz approach, we identify the existence of the Yu-Shiba-Rusinov-like state previously found in the non-Hermitian single-channel Kondo model. Then, based on non-Hermitian numerical renormalization group calculations, we reveal clear numerical signatures of the Yu-Shiba-Rusinov state emerging in the relatively strong non-Hermiticity regime of the \(\mathcal{PT}\)-asymmetric model. Furthermore, our boundary conformal field theory, which is found to be applicable for the \(\mathcal{PT}\)-symmetric model, uncovers an anomalous temperature dependence of the Kondo conductance, which is beyond conventional Hermitian Kondo systems. 
	\end{abstract}
	
	\maketitle
	
	\section{Introduction}
	Non-Hermitian quantum models, which physically arise from the loss or gain via coupling to an external environment, are well known to host a wealth of quantum phenomena absent in the Hermitian counterparts, e.g., the non Hermitian skin effect~\cite{zhao_two-dimensional_2025,yoshida_non-hermitian_2024,song_non-hermitian_2019-1,okuma_topological_2020,kawabata_entanglement_2023,li_critical_2020,PhysRevLett.124.250402,manna2023,PhysRevLett.124.056802}, non-Hermitian topology~\cite{ochkan_non-hermitian_2024,yao_edge_2018,bergholtz_exceptional_2021,yang_homotopy_2024,song_non-hermitian_2019,hamanaka_non-hermitian_2024,kawabata_symmetry_2019,gong_topological_2018,2zx3-rs57,ljvt-w6hw}, exceptional points~\cite{lee_exceptional_2022,ding_non-hermitian_2022,tang_exceptional_2020,miri_exceptional_2019,PhysRevLett.134.153601} and unexpected dynamics~\cite{PhysRevLett.134.180405,w377-f9mx,5ksl-tjjm,llbb-pcgk,z9m1-3mwb,PhysRevLett.133.070801,PhysRevLett.128.120401}. These findings open a new avenue in which many key concepts or tools for closed systems find new applications.  However, most existing works investigate non-Hermitian effects in non-interacting or weakly interacting systems, where quasi-particle descriptions are well-defined and applicable. Although there has been some studies on correlated non-Hermitian systems, they mainly focus on specific solvable models~\cite{bhpz-17d2,PhysRevLett.132.086502,PhysRevLett.133.193001,PhysRevB.111.L201106,PhysRevB.111.224407,nakagawa_non-hermitian_2018} as well as their ground state properties. The question of how the non-Hermiticity would affect correlated systems without well-defined particles has not been addressed yet. Particularly, it remains unclear whether or not any universal collective behavior would emerge in such cases.
	
	A typical and important example that violates the quasi-particle description is the non-Fermi liquid (NFL). It is one of the core concepts in modern condensed matter physics, describing a novel class of emergent phenomena beyond the Landau Fermi liquid paradigm. In strong-correlated systems where the conventional quasi-particle picture breaks down, there could emerge exotic collective behaviors, such as the unconventional scaling of thermodynamic quantities versus temperatures. A thorough understanding of these correlated phenomena is of key importance, as they are fundamentally related to the high-temperature superconductors~\cite{varma_phenomenology_1989,lee_doping_2006,zhang_high-temperature_2024,jiang_interplay_2023,yuan_scaling_2022,cao_strange_2020,chudnovskiy_superconductor-insulator_2022}, quantum spin liquids~\cite{anderson1973resonating,balents2010spin,RevModPhys.89.025003,Savary_2017,broholm2020science,PhysRevB.105.195156}, topological orders~\cite{wen1990topological,wen1990ground,levin2006detecting,chen2010local}, etc. On the other hand, the mechanism driving the non-Fermi liquid behavior can be various. It could arise from strong coupling between electrons and critical bosons, thus could naturally take place around quantum criticalities~\cite{lohneysen2007fermi,millis1993effect,moriya2012spin}. Besides, it can also be driven by the proximity to quantum spin liquids~\cite{anderson1987resonating,kalmeyer1987equivalence,kivelson1987topology,oike2017anomalous,PhysRevB.59.5341,PhysRevB.110.L121110}, which indicates that the frustration and correlation are essential for its generation. 
	
	Experimental signatures of NFL have been observed in various correlated systems, including heavy fermion materials~\cite{andres_4f-virtual-bound-state_1975,si_heavy_2010,stewart_non-fermi-liquid_2001,lohneysen_non-fermi-liquid_1994,coleman_theories_1999,custers_break-up_2003,PhysRevLett.127.026401,PhysRevLett.114.177202,PhysRevLett.134.116605,chang_mobius_2017}, high-$T_c$ superconductors~\cite{varma_phenomenology_1989,lee_doping_2006,zhang_high-temperature_2024,jiang_interplay_2023,yuan_scaling_2022,cao_strange_2020,chudnovskiy_superconductor-insulator_2022}, etc. Among them, the multichannel Kondo effect has attracted enormous interest, as it serves as prototypical platform for the experimental control over the NFL behaviors~\cite{iftikhar_tunable_2018,iftikhar_two-channel_2015,potok_observation_2007,keller_universal_2015,wen2002quantum,seaman_evidence_1991}. To realize a multichannel Kondo system that drives the NFL fixed point, exact channel symmetry should be achieved. Although this has been experimentally demonstrated by fine-tuning in nanostructures~\cite{iftikhar_tunable_2018,iftikhar_two-channel_2015}, another remarkable proposal is to utilize the topological Kondo effect. Based on multi-junction setup where Majorana fermions tunnel into normal leads, B\'{e}ri and Cooper demonstrate a feasible and controllable way to realize the multi-channel Kondo model with exact channel symmetry without the need of fine-tuning~\cite{beri_topological_2012}. Considering that the multichannel Kondo model hosts non-Fermi liquid fixed point, which exhibits fractionalized degrees of freedom contributing to a fractional impurity entropy~\cite{PhysRevLett.128.146803,PhysRevLett.129.227703,PhysRevLett.130.146201}, it is intriguing and worthwhile to ask would any new collective physics be emerging when non-Hermiticity further comes into play? 
	
	Despite the multichannel feature, the non-Hermitian single channel Kondo problem has been proposed and studied in previous works ~\cite{PhysRevB.111.L201106,nakagawa_non-hermitian_2018}. Ref.~\cite{nakagawa_non-hermitian_2018} carried out a perturbative renormalization group calculations and found that there occur two fixed points, respectively in the weak- and strong-coupling regime. Due to the non-Hermitian effects, there occurs a circling RG flow around the weak-coupling fixed point, driving the system towards the local moment phase. Moreover, it is argued that the strong-coupling regime describes the Kondo phase, in analogy with conventional Kondo physics of the Hermitian counterpart. A more recent study~\cite{PhysRevB.111.L201106} then performed a more careful solution of the Bethe ansatz solution. Remarkably, it reveals a more interesting phase, where the impurity is screened by effective extra degrees of freedom arising from the non-Hermitian effect, forming a single-mode bound state. Owing to its similarity with the Yu-Shiba-Rusinov (YSR) impurity states in conventional $s$-wave superconductors, it is dubbed the YSR phase. These results indicate that the non-Hermiticity brings about a richer phase diagram than expected, which is worth a detailed and comprehensive study by non-perturbative methods. 
	
	In this work, we perform a comprehensive, multi-method investigation of the non-Hermitian Kondo model, with a particular focus on the multichannel cases. Our aim here is threefold. First, although the multi-junction setup~\cite{beri_topological_2012} can be used to realize the conventional Hermitian multichannel Kondo model, it remains unclear how to systematically introduce the non-Hermitian effects into the system. Thus, we propose a realistic setup and demonstrate a step-by-step realization scheme. Second, we use a number of methods to thoroughly study the non-Hermitian multichannel Kondo (NHMCK) model in order to achieve a more comprehensive understanding beyond perturbative perspectives, involving the perturbative renormalization group (RG), Bethe ansatz, non-Hermitian numerical renormalization group (NHNRG), non-Hermitian Kubo formula, as well as the boundary conformal field theory (BCFT). Particularly, we clarify the role of \(\mathcal{PT}\) symmetry by considering both the \(\mathcal{PT}\)-symmetric and \(\mathcal{PT}\)-asymmetric NHMCK. Different features in terms of the low-energy spectrum and the thermodynamic quantities are found for the two cases. For the \(\mathcal{PT}\)-asymmetric NHMCK, using the NHNRG method, originally proposed by Ref.\cite{19td-1k9s}, we reveal numerical evidenve for the YSR phase predicted by Ref.~\cite{PhysRevB.111.L201106}. For the \(\mathcal{PT}\)-symmetric model, we combine the NHNRG with an additional analytic proof and show that the low-energy physics in this case is well within the BCFT descriptions. Third, we go beyond previous work and explore the nontrivial transport behaviors generated by the NFL and the non-Hermitian physics. We find that both the weak-coupling and the strong-coupling fixed point exhibit unconventional Kondo conductance different from the Hermitian systems.  Around the weak-coupling FP, the impurity moment decouples from the conduction electrons even for antiferromagnetic Kondo couplings, leading to a universal low-temperature conductance, $\sim1/\mathrm{ln}^2(T/T_K)$. Although the conductance is similar to that of the underscreened Kondo models~\cite{posazhennikova_anomalous_2005,coleman_singular_2003,mehta_regular_2005}, it is driven by a completely different dissipation mechanism--the nH-induced decoupling.  Remarkably, around the strong-coupling fixed point, the BCFT analysis predicts an anomalous deviation in the Kondo conductance with raising temperature.  Such a non-Hermiticity enriched anomaly stems from an emergent BCFT with non-Hermitian boundary operators, which is unique to systems exhibiting both non-Hermiticity and non-Fermi liquid behaviors.
	
	Last, we mention that this present work is a companion to another work by us, i.e., Ref.~\cite{Letter}. More concise results with more clear pictures are discussed in Ref.~\cite{Letter}, while the technical aspects in terms of methods and more detailed discussions are included in this manuscript. 
	
	\section{summary of results}
    In this section, we will outline the content in the remaining part of the work and our main results. First, we will introduce the non-Hermitian multichannel Kondo Hamiltonian and demonstrate a step-by-step realization scheme of this model, as presented in Sec.~\ref{sec2}. Then,  in Sec.~\ref{prg}, we perform a perturbative RG analysis, which suggests two possible phases corresponding to the weak- and strong-coupling fixed points. They are referred to as the  weak-coupling local moment phase and the strong coupling Kondo phase , respectively.   In Sec.~\ref{bas}, we present the Bethe ansatz calculations that reveal the existence of an intermediate YSR-like phase, which is beyond the perturbative RG analysis. Then, to better confirm these phases, we carry out a NHNRG study of the model. In Sec.~\ref{nrg}, we first introduce the technical details of the implementation of the NHNRG method and then utilize it to investigate the \(\mathcal{PT}\)-asymmetric NHMCK model.  Our focus in Sec.~\ref{ptm} is on the pseudo-Hermitian model with \(\mathcal{PT}\) symmetry. We first show the NHNRG results for the low-energy spectrum and the thermodynamic quantities, which justify the validity of the strong-coupling Kondo phase predicted by the perturbative RG. Then, we outline a more rigorous proof showing that the BCFT approach still applies for this model with \(\mathcal{PT}\) symmetry. In Sec.~\ref{trans}, we explore the transport properties associated with both fixed points. For the weak-coupling fixed point, we present the calculation procedures based on the non-Hermitian Kubo formula, which is well applicable for the perturbative regime. Then, for the strong-coupling fixed point, we exploit the BCFT, and demonstrate the anomalous Kondo conductance generated by the interplay between non-Hermiticity and non-Fermi liquid physics. More discussions and conclusions are included in Sec.~\ref{conc}.

	\begin{table*}[htbp]
		\centering
		\begin{tabular}{clll}
			\hline
			\multirow{2}{*}{} & \hspace{3mm} weak-coupling & \hspace{1mm} intermediate & strong-coupling \\
			& local-moment phase & YSR-like phase & \hspace{1mm} Kondo phase \\
			\hline
			\hline
			Perturbative RG & \hspace{10mm} $\checkmark,\ {\color{blue}\boldsymbol{C}}$ & \hspace{5mm} $\bm{\times}$ & \hspace{4mm} $\checkmark $ \\
			\hline
			Bethe ansatz & \hspace{10mm} $\checkmark$ & \hspace{5mm} $\checkmark$ & \hspace{4mm} $\checkmark$ \\
			\hline
			NHNRG & \hspace{10mm} $\checkmark,\ {\color{red}\boldsymbol{T}}$ & \hspace{5mm} $\checkmark,\ {\color{red}\boldsymbol{T}}$ & \hspace{4mm} $\checkmark,\ \color{red}\boldsymbol{T}$ \\
			\hline
			BCFT & \hspace{10mm} $\bm{\circ}$ & \hspace{5mm} $\bm{\times}$ & \hspace{4mm} $\checkmark^*,\ {\color{blue}\boldsymbol{C}}^*,\ {\color{red}\boldsymbol{T}}^*$ \\
			\hline
			NH Kubo formula &\hspace{10mm} $\checkmark,\ {\color{blue}\boldsymbol{C}}$ & \hspace{5mm} $\bm{\times}$ & \hspace{4mm} $\bm{\times} $ \\
			\hline
		\end{tabular}
		\caption{Synthesis table showing the applicability of different methods to the three distinct phases of the non-Hermitian multichannel Kondo model. The marker ``${\checkmark}$'' represents for ``applicable",``$\bm{\times}$'' denotes ``inapplicable", and ``$\bm{\circ}$'' means that the method could be applicable in principle but not studied in our work. The blue symbol ``${\color{blue}\boldsymbol{C}}$" means that the Kondo conductance behaviors are derived, and the red  symbol ``${\color{red}\boldsymbol{T}}$"  suggests that the thermodynamical properties are obtained. The superscript {*} indicates that the method is only applicable in the IR limit within the $\mathcal{PT}$-symmetric regime.}
		\label{tab1}
	\end{table*}

For clarity, we now summarize our main results, focusing on the three possible phases, i.e., the weak-coupling local moment phase, the intermediate YSR-like phase, and the strong coupling Kondo phase, and how they could be revealed by different analytical and numerical approaches.

First, the perturbative RG analysis up to 2-loop order reveals two fixed points, which are the local-moment fixed point and the strong-coupling Kondo fixed point, as mentioned above. Based on the RG flow of Kondo coupling, we further adopt the non-Hermitian Kubo formula to calculate the conductance around the weak-coupling fixed point.  However, we note that the perturbative approach cannot yield useful information about the intermediate YSR-like phase. Meanwhile, the predictions about the strong-coupling fixed point needs to be further confirmed since it is essentially beyond the perturbation theory.

Second, by means of rigorously solving the Bethe ansatz equations of the non-Hermitian multichannel Kondo model, we identify all the three distinct phases, controlled by a non-Hermiticity parameter $\theta$. The intermediate YSR-like phase is characterized by a single-mode bound state that contributes to the impurity screening, which is beyond the reach of the perturbative RG analysis.

Third, using  the NHNRG method, we further calculate  the thermodynamic properties in both the UV and IR regimes with varying $\theta$. From the obtained results including the impurity entropy and the specific heat, we not only arrive at the weak-coupling and strong-coupling phases, but also uncover numerical signatures of the YSR-like bound state. These results are in support of the Bethe ansatz solutions, indicating a richer phase diagram beyond the scope of Hermitian Kondo models.

Fourth, we analyze the quantum transport properties in the weak-coupling local-moment phase and the strong-coupling Kondo phase for both the UV and IR limits.   The conductance in the weak-coupling regime is obtained perturbatively from a non-Hermitian (NH) extension of the Kubo formula. Since this NH Kubo formula cannot be applied to non-perturbative regimes, we restrict ourselves to $\mathcal{PT}$-symmetric models and utilize the non-perturbative BCFT techniques to extract the IR behavior of the conductance in the strong-coupling Kondo phase. This approach also yields thermodynamic quantities for the strong-coupling Kondo phase, consistent with the results obtained by the NHNRG approach.

All the above results are concisely demonstrated by Table.~\ref{tab1}, which contains a compact “cross-validation” summary of the three phases from a multi-method perspective.

	\section{Model Hamiltonian and possible realization setup}\label{sec2}
	The non-Hermitian multichannel Kondo problems constitute a prototypical model which supports both non-Fermi liquid and non-Hermitian physics. The standard model Hamiltonian of our main focus in this work is given by,
	\begin{equation}\label{eqb1}
		H=H_{\mathrm{lead}}+
		\sum^n_{k=1}\sum_{l=1,2,3}J^{\prime}_ls_lc^{\dagger}_{k,\alpha}\tau_{l,\alpha\beta}c_{k,\beta},
	\end{equation}
	where $s_l$ is the impurity spin operator, $\tau_{l}$ is the 2 by 2 Pauli matrices describing the spin degrees of freedom of conduction electrons, $c_{k,\alpha}$ denotes the conduction electrons of ``spin" $\alpha$ and channel $k$, $H_{\mathrm{lead}}$ describes the conduction electrons, $J^{\prime}_l$ is a complex Kondo coupling with a nonzero imaginary component, with $l=1,2,3$ denoting $x,y,z$, and $J^{\prime}_l=J^{\prime}$ for the isotropic coupling case. Note that $J^{\prime}_l$ is independent of the channel index $k$, thus Eq.\eqref{eqb1} has channel symmetry, ensuring emergence of non-Fermi liquid physics.
	
	Although Eq.\eqref{eqb1} hosts rich physics, it is intriguing to ask how to realize the model, at least conceptually,  in condensed matter systems. In the following, we will demonstrate a two-step realization of the non-Hermitian multichannel Kondo model based on the quantum-dot-assisted tunneling junctions with Majorana fermions. 
	
\begin{figure}
		\includegraphics[width=\linewidth]{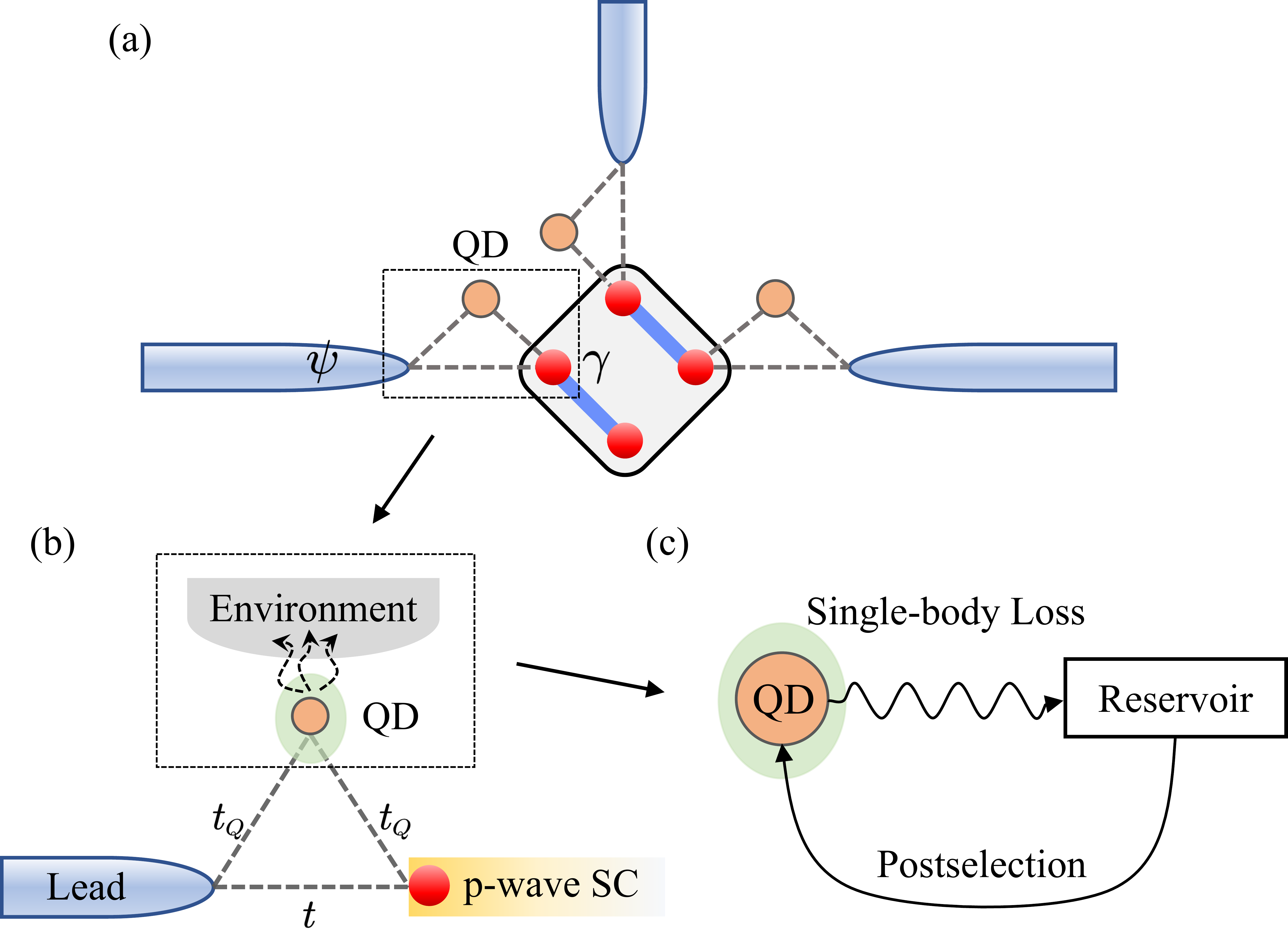}
		\caption{ The schematic plot of the realization setup of the NHMCK model. (a) The multi-junction nanostructure, where the Majorana modes (the red dot) on a superconducting island are connected to normal leads. The tunneling between the Majorana mode and the lead is assisted by a quantum dot (the orange dot). (b) A zoom-in plot of the Majorana-lead tunneling junction via the quantum dot. The quantum dot is further coupled to an environment. (c) The quantum dot has a single-body loss into the environment and monitored by a reservoir,  and a post-selection is implemented on the Lindbladian dynamics of the quantum dot.}
		\label{fig:setup}
	\end{figure}

	The setup is shown in Fig.\ref{fig:setup}(a). We consider a mesoscopic superconducting island that supports $M_{\mathrm{tot}}$ Majorana modes~\cite{beri_topological_2012,beri_majorana-klein_2013}, denoted by the Majorana operator $\gamma_{\alpha}$, and $\alpha=1,2,...M_{\mathrm{tot}}$. Then, $M$ out of the $M_{\mathrm{tot}}$ modes are coupled to normal leads with the assistance of a dissipative quantum dot (QD).  The island is connected to the ground by a capacitor, which contributes to a charging energy $H_c(N)=E_c(N-\frac{q}{e})^2$, where $N$ is the number of electrons on the island, and $q$ is the background charge determined by the voltage across the capacitor. Note that, without the dissipative QD, the setup is reduced to that proposed by B\'eri  and Cooper ~\cite{beri_topological_2012}, which realizes a Hermitian multichannel Kondo problem with exact channel symmetry.  
	
	The tunneling junction is plotted in more details in Fig.\ref{fig:setup}(b), which involves the dissipative QD. The latter consists of a QD coupled to environment. The QD-environment coupling system is further subjected to a monitoring-feedback device, which resets the particle loss events and ensures the particle conservation of the QD, as shown by Fig.\ref{fig:setup}(c).
	
	The quantum dot and environment form an isolated quantum system which evolves unitarily under the Hamiltonian.
	\begin{equation}\label{eqb2}
		H_{\mathrm{tot}}=H_{\mathrm{QD}}\otimes\mathbb{I}_{\mathrm{E}}+\mathbb{I}_{\mathrm{QD}}\otimes H_{\mathrm{E}}+H_{\text{int }},
	\end{equation}
	where $H_{\mathrm{QD}}  =\sum^{N_d}_{i=1}\epsilon_{d,i}d_{i}^{\dagger}d_{i}$  is the quantum dot Hamiltonian where $\epsilon_i$ is the energy level and $N_d$ the number of states.  $\mathbb{I}_{\mathrm{QD}} $ and $\mathbb{I}_{\mathrm{E}} $ are the identity operators defined in the Hilbert spaces of the system and the environment, respectively. The evolution of the total system density matrix then obeys unitary dynamics. However, the (reduced) density matrix of the QD alone is generally non-unitary. Following the standard procedure of tracing out the environment under necessary approximations~\cite{daley_quantum_2014,breuer_theory_2002,breuer_colloquium_2016}, the non-unitary dynamics of the QD is obtained, leading to the Lindblad master equation,
	\begin{equation}\label{eqb3}
		\frac{\mathrm{d}\varrho(t)}{\mathrm{d}t}=-i\left[H^{\prime},\varrho(t)\right]+\sum_{i}\left[L_{i}\varrho(t)L_{i}^{\dagger}-\frac{1}{2}\left\{ L_{i}^{\dagger}L_{i},\varrho(t)\right\} \right],
	\end{equation}
	where $\varrho$  is the reduced density matrix of the QD, $H^{\prime}=H_{\mathrm{QD}} +H_{\mathrm{LS}}$ where $H_{\mathrm{LS}}$ is the Lamb-shift term taking the form of $H_{\mathrm{LS}}=\sum_i \Omega_id_{i}^{\dagger}d_{i}$, with $\Omega_i$ being a pesudo-potential for QD acquired from coupling to the environment. The jump operators $L_{i}$, which  originate from the coupling of the quantum dot operators to the bath, can be constructed through the following manner,~\cite{lidar2020lecturenotestheoryopen,zhan_2026_rapid,breuer_theory_2002,weimer_simulation_2021},
	\begin{equation}\label{eqb4}
		L_i=\int_{-\infty}^{\infty}f_i(s)\mathrm{e}^{iH_\mathrm{tot}s} A_i\mathrm{e}^{-iH_\mathrm{tot}s}ds
	\end{equation}
	where $A_i$ denotes the coupling operators taken from the quantum dot, and $f_i(s)$ is the filter function. Thus, the quantum jump operator can be engineered via modulating the QD-environment coupling \cite{PhysRevB.105.195156,PhysRevB.109.235139}.
	
	In the following, we consider the single-body leakage into the bath, i.e. $A_i=d_{i}$, while more analysis of QD-environment coupling is included in Appendix~\ref{fen}.  Then, Eq.\eqref{eqb4} leads to $L_i=\sqrt{\varGamma} d_{i}$, where $\varGamma$ is the decay rate . Then the Lindblad master equation for the dissipative quantum dot is cast into,
	\begin{equation}\label{eqb5}
		\frac{\mathrm{d}\varrho(t)}{\mathrm{d}t}=-i\left[H^{\prime},\varrho(t)\right]+\sum_{i}\varGamma\left[d_{i}\varrho(t)d_{i}^{\dagger}-\frac{1}{2}\left\{ d_{i}^{\dagger}d_{i},\varrho(t)\right\} \right],
	\end{equation}
	Accordingly, the occupation number of the quantum dot can be calculated via Eq.\eqref{eqb5} as,
	\begin{equation}\label{eqb6}
		\sum_{i} \left\langle d_{i}^{\dagger}d_{i}\right\rangle_t = \sum_{i} \operatorname{Tr}\{\varrho(t)d_{i}^{\dagger}d_{i}\}\sim\exp(-\varGamma t),
	\end{equation}
	which decays with time due to the single-body leakage into the environment. 
	
	In order to realize a non-Hermitian system, the particle leakage event should be monitored by an attached empty reservoir as a particle detector~\cite{daley_quantum_2014,PhysRevB.111.125157} and then an immediate resetting needs to be implemented by the feedback equipment.  An alternative option is to take the record in the time period when no particle leakage is monitored. These procedures result in a postselection on the Lindbladian, leading to  an effective non-Hermitian Hamiltonian description of the QD, i.e.,
	\begin{equation}\label{eqb7}
		\frac{\mathrm{d}\varrho(t)}{\mathrm{d}t}=-i\sum_{k}\left[H_{\textrm{eff}},\varrho(t)\right]_+
	\end{equation}
	where $ \left[H,\bullet\right]_+=H \bullet+  \bullet H^{\dagger} $ is the non-Hermitian commutator. The effective Hamiltonian of QD reads as 
	\begin{equation}\label{eqb8}
		H_{\textrm{eff}}=\sum_{i}(\tilde\epsilon_{d,i}-i\frac{\varGamma}{2})d_{i}^{\dagger}d_{i},
	\end{equation} 
	where  $\tilde{\epsilon}_{d,i}$ denotes the Lamb-shifted QD energy level.  It is clear from above that the coupling to the environment and the postselection generate an dissipative QD with an imaginary chemical potential.  Note that the effective non-Hermitian Hamiltonian description can be also accessed through the Green's function point of view (see Appendix~\ref{fen}).
	
	We also mention in passing that, since under postselection the conservation of particle number (of the QD) is guaranteed, the time-dependent density matrix needs to be properly normalized for consistency~\cite{brody_mixed-state_2012}, i.e.,
	\begin{equation}\label{eqb9}
		\hat{\varrho}(t) = \frac{e^{-{ i}H_{\textrm{eff}}t} \varrho_0 
			e^{{ i}H_{\textrm{eff}}^{\dagger} t}} 
		{\text{Tr}( e^{-{ i}H_{\textrm{eff}}t} \varrho_0 
			e^{{ i}H_{\textrm{eff}}^{\dagger} t})}, 
	\end{equation}
	such that the particle number is obtained as,
	\begin{equation}\label{eqb10}
		\begin{aligned}
			n_t&=\sum_i\text{Tr}[\hat{\varrho}(t)d_{i}^{\dagger}d_{i}]=\sum_i\text{Tr}[\varrho(t)d_{i}^{\dagger}d_{i}]/\text{Tr}\varrho(t)\\
			&=\sum_i\text{Tr}[\varrho(0)d_{i}^{\dagger}d_{i}]/\text{Tr}\varrho(0)=n_0.
		\end{aligned}
	\end{equation}
	This normalized density matrix will be used later in our calculation of Kondo conductance, as will be shown in Sec.~\ref{trans}. 
	
	Then, we consider  the tunneling junction between the Majorana mode and the normal lead via the dissipative QD, as shown by Fig.\ref{fig:setup}(b). The junction Hamiltonian reads as, 
	\begin{equation}\label{eqb11}
		H_{\mathrm{junc}}= H_{\mathrm{QD,eff}}+H_{\mathrm{s-QD}}+ H_{\mathrm{s}}.
	\end{equation}
	It consists of the tunneling processes involving the QD, 
	\begin{equation}\label{eqb12}
		H_{\mathrm{s-QD}}=t_{Q}(\gamma+\psi^{\dagger})d+h.c.,
	\end{equation}
	where we have focused on the single-mode case with $N_d=1$ thus neglected the index $i$ for simplicity. The Majorana-lead tunneling is given by,
	\begin{equation}\label{eqb13}
		H_{\mathrm{s}}=(t\gamma\psi^{\dagger}+h.c.)+H_{\mathrm{lead}},
	\end{equation}
	where $\gamma$ denotes the Majorana mode,  $\psi$ represents for the conduction electron at the tunneling point, $t_Q$ and $t$ are the Majorana and QD tunneling coefficient, respectively. 
	
	We now show that the dissipative effect of the QD can induce a non-reciprocal tunneling process of the Majorana fermions.  This can be read off from the Schrödinger equation of the junction system, i.e., $H_{\mathrm{junc}}|\Psi\rangle=E|\Psi\rangle$, which is written as a coupled equation in the direct product space of the QD and the rest part. Tracing out the QD degrees of freedom then generates an effective junction Hamiltonian renormalized by the dissipative QD as, 
	\begin{equation}\label{eqb14}
		H_{\mathrm{junc,eff}}= H_{\mathrm{s}}+H_{\mathrm{s-QD}}(E-H_{\mathrm{QD,eff}})^{-1}H^{\mathrm{T}}_{\mathrm{s-QD}}
	\end{equation}
	Inserting of Eq.\eqref{eqb8}, we find that the original tunneling amplitude is renormalized in a nontrivial way. The tunneling from the Majorana to the lead ($t$) and the reversed tunneling $(t^{\star})$ are respectively modified as (see Appendix~\ref{tunneling}),  
	\begin{equation}\label{eqb15}
		t^{\prime}_+=t+\frac{|t_{\mathrm{Q}}|^2}{E-\tilde{\varepsilon}_d+i\Gamma/2},~~~~ t^{\prime}_-=t^{\star}+\frac{|t_{\mathrm{Q}}|^2}{E-\tilde{\varepsilon}_d+i\Gamma/2},
	\end{equation}
	Clearly,  the tunneling amplitude becomes non-Hermitian because  $t^{\prime\star}_+\neq t^{\prime}_-$. In the gauge where $t$ is real, both $t^{\prime}_{+}$ and $t^{\prime}_{-}$ are reduced to $ t^{\prime}=t^{\prime}_r+it^{\prime}_i$, where
	\begin{equation}\label{eqb16}
		t^{\prime}_r=t-\frac{|t_Q|^2\tilde{\varepsilon}_d}{\tilde{\varepsilon}^2_d+(\Gamma/2)^2}, ~~ t^{\prime}_i=\frac{|t_Q|^2\Gamma/2}{\tilde{\varepsilon}^2_d+(\Gamma/2)^2}.
	\end{equation}
	Note that the QD is assumed to have a higher energy scale than the junction system, i.e. $\tilde{\varepsilon}_d\gg E$. From above we know that  the dissipative QD endows the tunneling process of Majorana fermions with non-Hermiticity. This offers a convenient stereotypical generalization of the B\'eri -Cooper device~\cite{beri_topological_2012}, and realizes a non-Hermitian version of topological Kondo effect. The setup will support both the non-Hermiticity and the multichannel  feature with exact channel symmetry, as will be demonstrated in the following.  
	
	Consider the multi-lead junction as shown by Fig.\ref{fig:setup}(a),  for $E_c\gg|t_{\alpha}|$, the Schrieffer-Wolff transformation can be performed to eliminate the high-energy Bogoliubov particle states, leading to the following non-Hermitian effective Hamiltonian~\cite{beri_topological_2012},
	\begin{equation}\label{eqb17}
		H=H_{\mathrm{lead}}+\frac{1}{2}\sum_{\alpha\neq\beta}\lambda_{\alpha\beta}\gamma_{\alpha}\gamma_{\beta}\psi^{\dagger}_{\beta}\psi_{\alpha}=H_{\mathrm{lead}}+\sum_{l} J_{l}s_{l}\cdot S_{l},
	\end{equation}
	where $H_{\mathrm{lead}}$ describes the free conduction electrons on leads, $\psi_{\alpha}$ denotes the electron operator of the $\alpha$-th lead at the tunneling point, $s_{l}=-i\epsilon_{l\alpha\beta}\gamma_{\alpha}\gamma_{\beta}$ and $S_{l}=-i\epsilon_{l\alpha\beta}\psi^{\dagger}_{\beta}\psi_{\alpha}$, with $\epsilon_{l\alpha\beta}$ being the antisymmetric Levi-Civita tensor.  The effective couplings between the Majorana fermions and the lead fermions are obtained as, 
	\begin{equation}\label{eqb18}
		\lambda_{\alpha\beta}=[\frac{1}{E_c(N+1)-E_c(N)}+\frac{1}{E_c(N-1)-E_c(N)}]t_{\alpha}t_{\beta},
	\end{equation}
	and $J_{\lambda}=|\epsilon_{\lambda\alpha\beta}|\lambda_{\alpha\beta}$ are both complex, and $\lambda_{\alpha\beta}$ can be written into $\lambda_{\alpha\beta}=\lambda_r+i\lambda_i$ for the isotropic case where $t_{\alpha}=t$ for $\alpha=1,2,...M$.
	
	Notably, for different $M_{\mathrm{tot}}$ and $M$, Eq.\eqref{eqb17} can be further mapped to multichannel Kondo models of various underlying symmetries. For the minimal $M_{\mathrm{tot}}=4$ case~\cite{beri_topological_2012} with $t_{\alpha}=t$, Eq.\eqref{eqb17} is exactly reduced to Eq.\eqref{eqb1} with isotropic $J^{\prime}$, where $c_{k,\alpha}$ denotes the ``rotated" electron operator of the leads. Then, for $M=3$, the electron operators from the three leads join to form a spin-1 generator. Meanwhile, the Majorana modes encode an $\mathrm{SU}(2)$ Lie algebra, thus forming an effective pseudospin represented by spin-1/2 Pauli matrices.  Using the Kac-Moody algebra duality between $\mathrm{SO}(3)_1$ and $\mathrm{SU}(2)_4$  current~\cite{fabrizio_toulouse_1994}, this is mapped to a four-channel Kondo  model.  For $M=4$, the model has an $\mathrm{SO}(4)$ symmetry which contains two sets of  $\mathrm{SU}(2)$, giving rise to an $\mathrm{SU}(2)_2$ two-channel Kondo problem.  Other dualities also exist for larger $M$ ($M<M_{\mathrm{tot}}$), e.g. $\mathrm{SO}(6)\sim \mathrm{SU}(4)$, leading to multichannel $\mathrm{SU}(N)$ Kondo problems for general $M$~\cite{li_multichannel_2023,zazunov_transport_2014,altland_bethe_2014}. In addition, when $t_{\alpha}$ depends on $\alpha$, anisotropic models with different $J^{\prime}_l$ can also be achieved. Interestingly, tuning the Majorana tunneling to $t_3=t$, $t_1=t^{\star}_2=te^{i\theta}$, with $\theta$ being an angle that parametrizes the non-Hermiticity,  an anisotropic NHMCK model is realized with $J^{\prime}_3=J^{\prime}$, $J^{\prime}_1=J^{\prime\star}_2=J^{\prime}e^{i\theta}$, which is \(\mathcal{PT}\) symmetric. This is because under the spatial inversion operation $\mathcal{P}$, $J^{\prime}_1$ and $J^{\prime}_2$  are interchanged, and under the time-reversal operation $\mathcal{T}$, $i\rightarrow-i$, leaving the Kondo couplings invariant under the combined $\mathcal{PT}$. We describe the Hamiltonian invariant under $\mathcal{PT}$ as the $\mathcal{PT}$ symmetric model, and term it $\mathcal{PT}$ asymmetric otherwise. Thus, the setup shown in Fig.\ref{fig:setup}(a) provides a controllable platform for exploring non-Hermitian multichannel Kondo physics with tunable $\mathcal{PT}$ symmetry.
	
	Alternatively, we mention that the non-Hermitian multichannel Kondo model could also be experimentally realized using  cold-atom implementations. Recent studies have shown that both the localized magnetic moments and the multicomponent fermionic baths can be generated by engineering of the multi-orbital ultracold-atoms~\cite{Gorshkov_2010,PhysRevLett.120.143601}. Besides, the dissipative effect can be effectively introduced by ultracold alkaline-earth-like atom with inelastic scattering~\cite{PhysRevLett.124.203201,nakagawa_non-hermitian_2018} or ultracold atoms confined in optical lattice with fine-tuning~\cite{gong_topological_2018,doi:10.1126/science.1155309,10.1093/ptep/ptaa094}. Available monitoring installation can also be integrated in ultracold atoms~\cite{doi:10.1126/science.1155309,10.1093/ptep/ptaa094,Bouganne_2020,skin_2025,universal_2025,Kondozeno_2025}. 
	
	\section{Perturbative renormalization group results}\label{prg}
	In this section, we perform a perturbative renormalization group (RG) study of Eq.\eqref{eqb1}.Before studying the non-Hermitian effect, we will first revisit the Hermitian case by turning off the imaginary part of $J^{\prime}_l$. Taking the $M=3$ case as an example and following Anderson's poor man's scaling~\cite{anderson_poor_1970}, the RG flow equation, which is in terms of the $\beta$ function of $g_l=\nu_0J^{\prime}_l$ with $\nu_0$ being the density of states of conduction electrons near the Fermi surface, is derived as, 
	\begin{equation}\label{eqb19}
		\beta(g_{\alpha})=\frac{dg_{\alpha}}{db}=-\frac{dg_{\alpha}}{d\ln\Lambda}=g_{\alpha+1}g_{\alpha+2} ,\quad g_{\alpha+3}=g_{\alpha},
	\end{equation}
	where $\Lambda=\Lambda_0 e^{-b}$ defines the energy scale, with $\Lambda_0$ being the bare bandwidth of conduction electrons, and $b$ is the RG scaling parameter. 
	
	From Eq.\eqref{eqb19} we further obtain, 
	\begin{equation}\label{eqb20}
		\frac{dg_{\alpha}}{dg_{\alpha+1}}=\frac{g_{\alpha+1}}{g_\alpha}\Rightarrow g_{\alpha+1}^2-g_{\alpha}^2=C.
	\end{equation}
	This suggests that the anisotropy of tunneling junctions is irrelevant. Therefore we take $g_\alpha\approx g_{\alpha+1}\approx g_{\alpha+2}=g$, and Eq.\eqref{eqb19} reduces to
	\begin{equation}\label{eqb21}
		\beta(g)=\frac{dg}{db}=g_{}^{2}.
	\end{equation}
	From the equation above, we then obtain the Kondo temperature $T_K\sim \Lambda_0 e^{\frac{-1}{\overline{g}}}$ as $T\rightarrow T_K , g\rightarrow \infty$. These are known results for conventional Hermitian Kondo problems. 
	
	Then, we consider the general non-Hermitian $n$-channel Kondo model with  isotropic coupling, i.e., $g_l=g$ but is complex with $g=g_r+ig_i=ge^{i\theta}$. The RG flow equation to the third-order perturbation is obtained as,
	\begin{equation}\label{eqb22}
		\frac{\mathrm{d}g}{\mathrm{d}b}=g^2-ng^3/2,
	\end{equation}
	which gives rise to,
	\begin{align}\label{eqb23}
		\beta(g_r)=\frac{dg_r}{db}=g_r^2-g_i^2-\frac{n}{2}(g_r^3-3g_rg_i^2),\\
		\beta(g_i)=\frac{dg_i}{db}=2g_rg_i-\frac{n}{2}(3g_r^2g_i-g_i^3).
	\end{align}
	The equations lead to a circling flow in the $g_r,g_i$ plane, as shown by Fig.3(b) of Ref.~\cite{Letter}.  Two fixed points are found at $(\frac{2}{n},0)$ and $(0,0)$, both lying on the real axis. $(0,0)$ corresponds to the weak-coupling fixed point and $(\frac{2}{n},0)$ corresponds to the strong-coupling fixed point. The two fixed points indicate two phases, dubbed the  strong-coupling phase and weak-coupling phase respectively, which are separated by a critical line (see Appendix~\ref{ktrg}).  
	
	For a fixed $g_r$ and an increasing $g_i$ with gradually enlarged non-Hermiticity, the system first flows to the strong-coupling phase for small $\theta$ and then flows to the weak-coupling phase for large $\theta$. Since the perturbative RG method is only expected to be valid and reliable for weak-coupling phase, it is intriguing to ask whether more exotic phases could occur in the strong or intermediate coupling regime.  Notably, recent Bethe ansatz studies have clearly implied that there indeed emerges an extra phase in the intermediate region for intermediate $\theta$~\cite{PhysRevB.111.L201106}, where additional degrees of freedom screen the local moment. This is termed as the Yu-Shiba-Rusinov (YSR) phase owing to its similarity with impurity states in superconductors~\cite{kattel2025multichannelkondoeffectsuperconducting,kattel2025thermodynamicssplithilbertspace}. Since this emerging new phase is missed by the perturbative RG analysis, it is obvious that more careful non-perturbative approaches are indispensable to achieve a better and comprehensive understanding of the non-Hermitian Kondo effect.

	\section{Bethe ansatz solution}\label{bas}
	In this section, we solve the general $\mathcal{PT}$ asymmetric NHMCK using the Bethe ansatz method, which is still applicable in the non-Hermitian setting, as previously studied in Ref.~\cite{PhysRevB.111.L201106,PhysRevB.111.224407}. 
	
	We consider the $n$-channel isotropic model in  Eq.\eqref{eqb1}, whose low-energy effective Hamiltonian reads as,
	\begin{equation}\label{eqb25}
		\begin{split}
			\hat{H}&=-iv_{f}\sum_{k,\alpha}\int dx\psi_{k,\alpha}^{\dagger}(x)\partial_{x}\psi_{k,\alpha}(x)\\
			&+2J\sum_{k,\alpha,\beta}\psi_{k,\alpha}^{\dagger}(0)\vec{\tau}_{\alpha\beta}\psi_{k,\beta}(0)\cdot\vec{s},
		\end{split}
	\end{equation}
	where $v_{f}$ denotes the Fermi velocity, $\psi_{k,\alpha}^{\dagger}(x),\psi_{k,\alpha}(x)$ are the creation and annihilation operators of conduction electrons (after the linearization of the spectrum around Fermi surface and the mapping to 1D model~\cite{affleck_current_1990,affleck_kondo_1991,affleck_critical_1991,affleck_exact_1993,ludwig_exact_1994,andrei_solution_1983,andrei_solution_1984}) defined at coordinate $x$ with spin $\alpha$ and channel $k$. $J$ is the bare complex coupling.
	The corresponding Bethe ansatz equations are obtained as,
	\begin{equation}\label{eqb26}
		\begin{gathered}
			e^{ik_jL}=\prod_{\alpha=1}^{M}\frac{\lambda_\alpha+i\frac{n}{2}}{\lambda_\alpha-i\frac{n}{2}}\\
			\left(\frac{\lambda_\alpha+in/2}{\lambda_\alpha-in/2}\right)^N\frac{\lambda_\alpha+1/J+i/2}{\lambda_\alpha+1/J-i/2}=-\prod_{\beta=1}^M\frac{\lambda_\alpha-\lambda_\beta+i}{\lambda_\alpha-\lambda_\beta-i},
		\end{gathered}
	\end{equation}
	where $k_j$ denote the momenta and $\lambda_\alpha$ are the spin rapidities.
	
	For simplicity, we introduce the following notation $\frac{1}{J}=\widetilde{g}e^{-i\theta}$, so the Bethe ansatz equations become 
	\begin{equation}\label{eqb27}
		\begin{gathered}
			e^{ik_jL}=\prod_{\alpha=1}^{M}\frac{\lambda_\alpha+i\frac{n}{2}}{\lambda_\alpha-i\frac{n}{2}}\\
			\left(\frac{\lambda_\alpha+in/2}{\lambda_\alpha-in/2}\right)^N\frac{\lambda_\alpha+\widetilde{g}e^{-i\theta}+i/2}{\lambda_\alpha+\widetilde{g}e^{-i\theta}-i/2}=-\prod_{\beta=1}^M\frac{\lambda_\alpha-\lambda_\beta+i}{\lambda_\alpha-\lambda_\beta-i}.
		\end{gathered}
	\end{equation}
	Taking logarithm for both sides of the above equations, we obtain 
	\begin{equation}\label{eqb28}
		N\Theta(\frac{2\lambda_\alpha}{n})+\Theta(2(\lambda_\alpha+\widetilde{g}e^{-i\theta}))=\sum_{\beta=1}^M\Theta(\lambda_\alpha-\lambda_\beta)-2\pi I_\beta,
	\end{equation}
	where $i\Theta(z)=\ln \frac{i+z}{i-z}$.
	
	The non-Hermitian Kondo model is characterized by two RG-invariant parameters~\cite{PhysRevB.111.L201106}, i.e., the Kondo energy scale $T_K$ and the non-Hermiticity parameter $\alpha$, given by,
	\begin{equation}\label{eqb29}
		T_K=2\nu e^{-\frac{\pi \cos(\alpha)}{\widetilde{g}}},\quad \alpha=\frac{\pi\sin{\theta}}{\widetilde{g}},
	\end{equation}
	where $\nu=\frac{N}{L}$ is the density of electrons. For simplicity, we set $\widetilde{g}=1$ in the following calculations.
	
	We first consider weak non-Hermiticity for $\sin{\theta}<\frac{1}{2}$. In this region, the density of $\lambda$, i.e., $\sigma(\lambda)$, is defined as $\sigma(\lambda)=\frac{dI}{d\lambda}$. In the continuum limit, Eq.\eqref{eqb28} can be written as
	\begin{equation}\label{eqb30}
		\begin{split}
			N\Theta(\frac{2\lambda_\alpha}{n})+\Theta(2(\lambda_\alpha+e^{-i\theta}))&=\int_C\sigma(\lambda^{\prime})\Theta(\lambda_\alpha-\lambda^{\prime})d\lambda^{\prime}\\
			&-2\pi I_\beta.
		\end{split}
	\end{equation}
	Taking derivative with respect to $\lambda$ on both sides of Eq.\eqref{eqb30}, we can obtain,
	\begin{equation}\label{eqb31}
		\sigma_0(\lambda)=f(\lambda)-\int_Ca(\lambda-\lambda^{\prime})\sigma_0(\lambda^{\prime})d\lambda^{\prime},
	\end{equation}
	where
	\begin{equation}\label{eqb32}
		f(\lambda)=\frac{2n}{\pi}\frac{N}{n^2+4\lambda^2}+\frac{2}{\pi}\frac{1}{1+4(\lambda+e^{-i\theta})^2},
	\end{equation}
	and
	\begin{equation}\label{eqb33}
		a(\lambda)=\frac{1}{\pi(1+\lambda^2)}.
	\end{equation}
	When considering the n-string solutions, the solution is given by 
	\begin{equation}\label{eqb34}
		\lambda^{n,m}_j=\lambda_j^n+\frac{i}{2}(n+1-2m)\quad m=1,2,\cdots n.
	\end{equation}
	In this case, the kernel is given by 
	\begin{equation}\label{eqb35}
		a_n(\lambda)=\frac{1}{\pi}\frac{n}{\lambda^2+n^2}+\frac{2}{\pi}\sum_{m=1}^{n-1}\frac{n-m}{\lambda^2+(n-m)^2}.
	\end{equation}
	The above equation can be solved in the Fourier space. Using $\mathcal{F}(A*B)=\hat{A}\hat{B}$  where $*$ denotes the convolution, one arrives at,
	\begin{equation}\label{eqb36}
		\hat{\sigma}_0(p)=\frac{1}{2}N \text{sech}(\frac{np}{2})+\frac{1}{2}e^{ip e^{-i\theta}}\text{sech}(\frac{p}{2}),
	\end{equation}
	which can be written in the $\lambda$ space via the inverse Fourier transform as, \footnote{For the simplicity, we omit $\pi/g$, the constant factor in equations.}
	\begin{equation}\label{eqb37}
		\sigma_0(\lambda)=\frac{N}{2n}\text{sech}(\frac{ \lambda}{n})+\frac{1}{2}\text{sech} (\lambda+e^{-i\theta}).
	\end{equation}
	Using  $\sigma_0(\lambda)$, we can calculate the ground magnetization when there is no special root involved. The result is 
	\begin{equation}\label{eqb38}
		S=\frac{N+1}{2}-\int \sigma_0(\lambda)d\lambda=\frac{N+1}{2}-(\frac{N}{2}+\frac{1}{2})=0,
	\end{equation}
	where we have used the identity, $\int \text{sech}(\frac{\pi x}{n})dx=n$. The total spin $S=0$ implies that the impurity is fully screened. 
	
	Now we discuss the charge and spin excitations. Since the charge degree is decoupled with the spin degrees of freedom, the excitation energy is given by,
	\begin{equation}\label{eqb39}
		\Delta E_c=\frac{2\pi}{L}\Delta n.
	\end{equation}
	In order to calculate the spin excitations, we insert a hole in the system. In the presence of the hole,  Eq.\eqref{eqb31} is modified into,
	\begin{equation}\label{eqb40}
		\sigma(\lambda)+\sigma^h(\lambda)=f(\lambda)-\int a(\lambda-\lambda^{\prime})\sigma(\lambda^{\prime})d\lambda^{\prime}.
	\end{equation}
	Given a single hole excitation, i.e.,
	\begin{equation}\label{eqb41}
		\sigma^h(\lambda)=\delta(\lambda-\lambda^{h}_1),
	\end{equation}
	we can solve the equation in Fourier space similarly to what was done above, i.e.,
	\begin{equation}\label{eqb42}
		\begin{split}
			\hat{\sigma}(p)&=\hat{\sigma}_0(p)+\Delta\hat{\sigma}_p \\
			&=\frac{1}{2}N \text{sech}(\frac{np}{2})+\frac{1}{2}e^{ip e^{-i\varphi}}\text{sech}(\frac{p}{2})-\frac{e^{-i\lambda^h_1p+n|p|/2}}{2\text{cosh}(\frac{np}{2})},
		\end{split}
	\end{equation}
	such that,
	\begin{equation}\label{eqb43}
		\begin{aligned}
			\Delta M&=\int \Delta \sigma(\lambda)d\lambda=\Delta\hat{\sigma}_p|_{p=0}=\frac{1}{2},\\
			\Delta \mathcal{E}_s&=\nu\int\Delta\sigma(\lambda)\Theta(2\lambda/n)d\lambda \\ &=2\nu \arctan (e^{ \pi \lambda_1^h/n})=E_h+i\Gamma_h.
		\end{aligned}
	\end{equation}
	The spin excitations carrying spin 1/2 are termed as the spinons. Note that the spinon energy is complex, which consists of both a real part $E_h$ and an imaginary part $\Gamma_h$. 
	
	Since we only consider the 1-string  solution, $\lambda^h=\mu_h+\frac{i}{N}\zeta(\mu_h)$, where the $\zeta(\mu)$ is the imaginary part of the rapidities with 1-string. Then the spinon energy can be readily calculated, the result is given by
	\begin{equation}\label{eqb44}
		\begin{aligned}
			E_h&=2\nu\arctan{e^{\pi\mu_h/n}},\\
			\Gamma_h&=\frac{\pi N}{nL\cosh{[\pi\mu_h/n]}}\zeta(\mu_h)\\ &=\frac{1}{L}\tanh^{-1}(\frac{2E_hT_K}{E_h^2+T_K^2}\sin{\alpha})\ll1,
		\end{aligned}
	\end{equation}
	where we have used $\arctan(x+y)\approx \arctan x+\frac{y}{1+x^2}$ for $x\gg y$.  The result implies that the imaginary part of spinon energy and thus the spinon decay rate scale as $\frac{1}{L}$.
	
\begin{figure*}[htb]
		\centering
		\includegraphics[width=\linewidth]{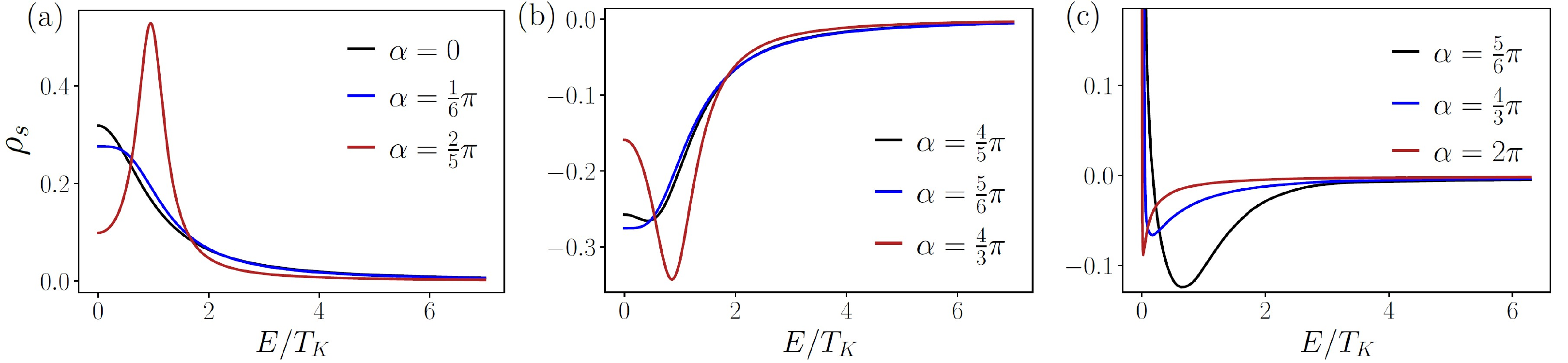}
		\caption{ Bethe ansatz calculations of multichannel non-Hermitian Kondo model without \(\mathcal{PT}\) symmetry. The calculated spinon DOS contributed by  the impurity for different $\alpha$ in the three phases, i.e., the Kondo phase, the bound-mode states in YSR-like phase, and the unscreened states (both YSR-like phase and unscreened phase). Similar results have been obtained for the single channel model in Ref.~\cite{PhysRevB.111.L201106}.}
		\label{fig:ba}
	\end{figure*}

	We can further obtain the density of state (DOS) of spinons contributed by the impurity, $\nu_s$, which is a good indicator for distinguishing different phases and states, namely,
	\begin{equation}\label{eqb45}
		\nu_s=\frac{\cos{\alpha}}{\pi T_K}\frac{1+(E/T_K)^2}{1+2(E/T_K)^2\cos(2\alpha)+(E/T_K)^4},
	\end{equation}
	which is a function of $\alpha$ and $\frac{E}{T_K}$. We then plot $ \nu_s$ for different $\alpha$, as shown in Fig.\ref{fig:ba}(a). When $0\leq\alpha\leq\pi/6$, the spinon DOS remains a pure Kondo behavior with a peak at $E=0$. When $\alpha\geq \pi/6$, the peak moves to $E/T_K=\sqrt{2\sin \alpha -1}$, until $\alpha$ approaches $\pi/2$. Meanwhile, the spinon DOS develops a delta peak, i.e.,  $\delta(E-T_K)/2$ at $\alpha=\pi/2$.  These results coherently indicate that the number of modes that screen the impurity decreases with $\alpha$, until reaching the case where the screening is contributed only by a single mode at $E/T_K=1$.
	
	Now we discuss the intermediate non-Hermiticity regime, i.e., $1/2<\sin{\theta}<3/2$, and will show that a new phase emerges, dubbed the YSR phase~\cite{PhysRevB.111.L201106}. This phase originates from the observation that  there is an extra impurity-string solution of Bethe equations apart from the dense roots $\mathcal{C}$, which describes a dynamically unstable bound mode with finite lifetime. Similar scenarios can be found in Hermitian cases with boundaries~\cite{PhysRevB.101.085133,PhysRevB.109.174416}. 
	
	The impurity string is given by, 
	\begin{equation}\label{eqb46}
		\lambda_{\mathrm{IS}}=-\frac{i}{2}-e^{-i\theta}=-\cos(\theta)+\frac{i}{2}(2\sin(\theta)-1).
	\end{equation}
	There are two possible states in this phase, which are distinguished by whether or not  the impurity-string is added to the dense solution $\mathcal{C}$. 
	
	The density of continuous root distribution in the Fourier space is 
	\begin{equation}\label{eqb47}
		\hat{\sigma}(p)=\frac{N}{2} \text{sech}(\frac{np}{2})+\frac{(e^p-1)\theta(-p)e^{-\frac{1}{2}p(1-2\sin\theta-2i\cos\theta)}}{e^{-|p|}+1}.
	\end{equation}
	Adding a hole will result in an impurity-unscreened state. Such a (unscreened) state will be discussed later. However, when adding an impurity string, the result becomes different.  The contribution of impurity string solution in Fourier space is,
	\begin{equation}\label{eqb48}
		\Delta \hat{\sigma}^{\mathrm{IS}}(p)= \begin{cases}-\frac{e^{-\frac{ p}{2}+i p e^{-i \theta}}}{e^{[p]}+1} ,&  3/2 > \sin \theta, \\ -\frac{\left(e^{2  p}-1\right) \theta(-p) e^{-\frac{3  p}{2}+i p e^{-i \theta}}}{e^{-|p|}+1} ,&  3 /2< \sin \theta.\end{cases}
	\end{equation}
	Note that the continuous root density is just the analytical continuation of the Kondo phase as in Eq.\eqref{eqb37}.  
	
	When the impurity string root is included,  similar to the Kondo phase, the impurity will be screened. However it is screened by a single-particle bound mode. Moreover, the change of local moment is computed to be,
	\begin{equation}\label{eqb49}
		\Delta M=1+  \Delta \hat{\sigma}^{\mathrm{IS}}(0)=\frac{1}{2},
	\end{equation}
	confirming the screening of the impurity. 
	
	In Fig.\ref{fig:ba}(b), we plot the spinon DOS contributed by the impurity in the YSR phase (the bound mode contribution is not shown. Note that the spinon DOS is an analytical continuation of Eq.\eqref{eqb45} plus the bound mode contribution, which is a delta function). As shown, it is found that the impurity contribution is negative in the  absence of impurity string solution. The negativity indicates that the spinons do not participate in the screening of the impurity, thus the screening is solely contributed by the single bound state, i.e., the YSR state.
	
	When the impurity solution is not included, the density of state in  $\mathcal{C}$ is given by, 
	\begin{equation}\label{eqb50}
		\sigma(\lambda)=\frac{N}{2n\cosh(\lambda/n)}+\frac{1}{2\pi}\left(\frac{1}{z}+\Psi(\frac{z}{2})-\Psi(\frac{z+1}{2})\right),
	\end{equation}
	where $z=i(\lambda-\lambda_{\mathrm{IS}})$ and $\Psi(z)$ is the digamma function. From the integral  of  the density of rapidities $\lambda$, one knows that this state has total spin  $S=1/2$, implying that the impurity is unscreened. The property of such a state is similar to the unscreened phase discussed below.
	
	Moreover, we note that in the region $1/2<\sin\theta<1$, the energy of the bound-mode state is lower than that of the unscreened state (in terms of the real part of energy), which means that the impurity 1-string is occupied in the ground state. In contrast, for $1<\sin\theta<3/2$, the impurity is unscreened, as this state is now more energetically favorable. A first-order phase transition would therefore take place at $\sin \theta=1$, in analogy with the singlet-doublet transition corresponding to the YSR state in $s$-wave superconductors~\cite{PhysRevB.105.174517,kattel2025multichannelkondoeffectsuperconducting,kattel2025thermodynamicssplithilbertspace}. 
	
	Last, we consider the strong non-Hermiticity region for $\sin\theta>\frac{3}{2}$. Similar to previous analysis, the impurity string solution is still present. However, we find the impurity string solution in this region has vanishing energy. This means that adding an impurity string will lead to  a degenerate state. In this phase, the density of state in $\mathcal{C}$ is given by 
	\begin{equation}\label{eqb51}
		\sigma(\lambda)=\frac{N}{2n\cosh(\lambda/n)}+\frac{1}{2\pi}\left(\frac{1}{i\lambda}+\Psi(\frac{i\lambda}{2})-\Psi(\frac{i\lambda+1}{2})\right).
	\end{equation}
	Correspondingly, the magnetization can be calculated via analytical continuation by using Eq.\eqref{eqb51}, i.e.,
	\begin{equation}\label{eqb52}
		M=\int\sigma(\lambda)d\lambda=\frac{N}{2},
	\end{equation}
	\begin{equation}\label{eqb53}
		S=\frac{N+1}{2}-M=\frac{1}{2}.
	\end{equation}
	It is therefore clear that the impurity is unscreened. 
	
	
	If one adds the impurity string root, the DOS in the unscreened phase is derived to be,
	\begin{equation}\label{eqb54}
		\begin{split}
			\nu_s(\mathcal{\mathcal{E}})&=[\frac{2\pi}{3\pi-2i\ln{\frac{\mathcal{E}}{T_0}}}+\psi^{(0)}(\frac{i\ln(\mathcal{E}/T_0)}{2\pi}-\frac{1}{4})\\
			&-\psi^{(0)}(\frac{i\ln(\mathcal{E}/T_0)}{2\pi}-\frac{3}{4})]/[2\pi^2 \mathcal{E}],
		\end{split}
	\end{equation}
	where $\psi^{(0)}$ is the digamma function and $\mathcal{E}$ is complex,  whose real part is $E$. As shown in Fig.\ref{fig:ba}(c),   depending on the energy, the impurity contribution can be either positive or negative, which could be attributed to a partial screening mechanism~\cite{PhysRevB.109.174416,Kattel_2024}.
	
	The above Bethe ansatz solution is obtained for the general non-Hermitian multichannel Kondo model with isotropic Kondo exchange couplings, thus is not applicable for the $\mathcal{PT}$ symmetric case. For the $\mathcal{PT}$ symmetric case (with  $J^{\prime}_3=J^{\prime}$, $J^{\prime}_1=J^{\prime\star}_2=J^{\prime}e^{i\theta}$), although we do not have access to the specific Bethe ansatz solution yet, we expect that only the contribution of the real part of impurity scattering could remain, since the imaginary part contributed by the conjugated $J^{\prime}_1$, $J^{\prime\star}_2$ pairs may cancel with each other, as will be implied by the NHNRG results below.  Here, for the $\mathcal{PT}$ asymmetric case, we  clearly see that the non-Hermitian effects can induce exotic phases absent in the Hermitian counterparts. Then, a urgent question to be answered is that how to obtain clear signatures of the emerging YSR phase, particularly regarding the thermodynamic quantities. We will further clarify this question in the next section using the NHNRG method.
	
	\section{non-Hermitian nrg approach}\label{nrg}
	
	\subsection{Numerical implementations}
	
	In this section, we demonstrate the details of the implementation of NHNRG, focusing on its difference with conventional NRG for Hermitian systems. 
    We note that a previous work by  P. C. Burke and A. K. Mitchell~\cite{19td-1k9s} developed a NHNRG approach and solved the non-Hermitian single-channel quantum impurity model.  Here, we further study the multichannel cases and focus on the thermodynamic quantities with varying the temperature (see Eq.~\eqref{eqb73} and~\eqref{eqb74} below). 
	
	The conventional NRG method involves several steps \cite{RevModPhys.80.395}. The first step is the logarithmical discretization. One reformulates the original impurity model using a one-dimensional (1D) energy representation for the bath electrons under a band width cutoff \(\pm D\). The conduction band can be discretized to a series of energy intervals \(D_n\) with a scaling parameter \(\Lambda\), i.e.,
	\begin{equation}\label{eqb55}
		x_n=\pm \Lambda^{-n},\ n=0,1,2,...,
	\end{equation}
	and the interval is \(d_n=\Lambda^{-n}(1-\Lambda^{-1})\). Within each energy interval, we can introduce a complete set of orthonormal wavefunctions. Then,  fermionic operators can be defined  in each interval. The second step is to map the discretized model into the Wilson chain. Through unitary transformations, the model is mapped into a semi-infinite 1D chain, where the impurity is placed at one end of the chain and coupled to the bath sites in a one-by-one manner through short-range terms. The last step is to iteratively diagonalize the Wilson chain. The Hamiltonian of the first \(N+1\) sites can be separated into two parts, i.e.,
	\begin{equation}
		H_{N+1} = \Lambda^{1/2} H_N + \Lambda^{N/2} h_{N+1},
		\label{eqb56}
	\end{equation}
	where \(H_N\) is the first \(N\) site Hamiltonian, and \(h_{N+1}\) represents short-range terms containing the \((N+1)\)-th site. In each iteration, we truncate the energy spectrum and dilate the kept eigenenergies, which gradually yields the low-energy physics. The success of this strategy is guaranteed by the energy scale separation of nearest sites in the Wilson chain, owing to the logarithmical discretization. The iteration diagonalization can then be stopped as long as a stable low-energy fixed point is reached. Such a diagonalization process also reveals information about the fixed points, as well as thermodynamic properties with varying energy scales. 
	
	To generalize the above NRG method to non-Hermitian systems, it is natural to adopt the biorthogonal basis.  For a non-Hermitian system \cite{ashida2020non}, both the left and the right eigenvectors need to be defined, which reads as,
	\begin{equation}
		\begin{array}{lll}
			H\ket{\psi_j}^R &= E_j \ket{\psi_j}^R,\quad \prescript{R}{}{\bra{\psi_j}} H^\dagger &= \prescript{R}{}{\bra{\psi_j}} E_j^*, \\
			H^\dagger \ket{\psi_j}^L &= E_j^* \ket{\psi_j}^L,\quad \prescript{L}{}{\bra{\psi_j}} H &= \prescript{L}{}{\bra{\psi_j}} E_j.
		\end{array}
		\label{eqb57}
	\end{equation}
	Although the two sets of wavefunctions are not orthogonal by themselves alone, they do form a biorthogonal basis that satisfies, \cite{brody2013biorthogonal}
	\begin{equation}
		\quad \prescript{L}{}{\braket{\psi_i|\psi_j}}^R=\delta_{ij}, \label{eqb58}
	\end{equation}
	where we have assumed a nondegenerate energy spectrum and an automatic normalization by taking
	\begin{equation}\label{eqb59}
		\ket{\psi_j}^R\to \frac{\ket{\psi_j}^R}{\sqrt{\prescript{L}{}{\braket{\psi_j|\psi_j}}^R}},\
		\prescript{L}{}{\bra{\psi_j}}\to \frac{{}^L\bra{\psi_j}}{\sqrt{\prescript{L}{}{\braket{\psi_j|\psi_j}}^R}^*}.
	\end{equation}
	The complex eigenvalues here indicate that the diagonalization is achieved via non-unitary transformation matrices, and in general we have
	\begin{gather}
		(U^L)^\dagger U^L \neq \mathbb{I},\ (U^R)^\dagger U^R \neq \mathbb{I},\label{eqb60} \\
		(U^L)^\dagger U^R = \mathbb{I}, \label{eqb61}
	\end{gather}
	where the columns of \(U^R\) and \(U^L\) are the right and left eigenvectors. We note that Eq.\eqref{eqb61} is invalid at the exceptional point where eigenvectors coalesce.
	
	We mention in passing that, if one imposes a physical space-time reflection symmetry by requiring,
	\begin{equation}\label{eqb62}
		H\mathcal{PT}=\mathcal{PT}H,
	\end{equation}
	then the eigenenergies can be proved to either be real (\(\mathcal{PT}\)-symmetric states) or appear in complex conjugate pairs (spontaneously \(\mathcal{PT}\)-broken states) \cite{bender2007making}. 
	
	From above, one observes that the differences between Hermitian and non-Hermitian NRG lie in their different ways of diagonalization. For non-Hermitian systems, we need to calculate both left and right eigenvectors for each iteration step, as clarified in Eq.\eqref{eqb57}. This generates more detailed issues regarding the non-unitary transformation, truncation, and the degeneracy, etc, which will be explained in the following.
	
	In the Hermitian framework, given a Hamiltonian \(H_N\) that has been diagonalized at the \(N\)-th step, we can construct the basis of the \(N+1\)-th step by the direct product,
	\begin{equation}\label{eqb63}
		\ket{ks}_{N+1} = \ket{k}_{N} \otimes \ket{s}_{N+1},
	\end{equation}
	where \(\ket{k}_{N}\) denote the kept states after truncation at the \(N\)-th step, \(\ket{s}_{N+1}\) represents the local basis of the \((N+1)\)-th site. Then, the Hamiltonian matrix for \(H_{N+1}\) of Eq.\eqref{eqb56} can be written as,
	\begin{equation}
		H_{N+1}(ks,k's') = \prescript{}{N+1}{\braket{ks|H_{N+1}|k's'}}_{N+1},
		\label{eqb64}
	\end{equation}
	where the notation \(H(m,n)\) stands for the $(m,n)$ element of the matrix \(H\). Diagonalization of the above matrix gives the eigenenergies and eigenstates of the Wilson chain of \((N+1)\)-sites. The eigenstates $\ket{w}_{N+1}$ are obtained via the unitary transformation,
	\begin{equation}\label{eqb65}
		\ket{w}_{N+1} = \sum_{ks} U_{N+1}(w,ks) \ket{ks}_{N+1},
	\end{equation}
	Then, after truncation, the set of eigenstates \(\{\ket{w}_{N+1}\}\) are turned to \(\{\ket{k}_{N+1}\}\), which will further be used to construct the basis for the next iteration. 
	
	
	During the $N$-th iterative diagonalization, the observable operators transform accordingly as,
	\begin{equation}\label{eqb66}
		A_N^{\{w\}} = U_N^\dagger A_N^{\{ks\}} U_N, 
	\end{equation}
	where the superscripts \(\{ks\}\), \(\{w\}\) denote the two different basis that are related by Eq.\eqref{eqb65}.  Then, the partition function and the expectation value of observables can be calculated by taking traces at each step, i.e.,
	\begin{equation}\label{eqb67}
		Z_N = {\rm Tr} e^{-\beta H_N} = \sum_{w} e^{-\beta E_{N,w}},
	\end{equation}
	and
	\begin{equation}\label{eqb68}
		\braket{A}_N = \frac{{\rm Tr} A_N e^{-\beta H_N}}{Z_N} = \frac{\sum_w \prescript{}{N}{\bra{w}} A^{\{w\}}_N \ket{w}_N e^{-\beta E_{N,w}}}{Z_N}.
	\end{equation}
	
	For non-Hermitian systems, the general mapping scheme to the 1D Wilson chain is still applicable. However, in contrast to the Hermitian case, the eigenenergies of the Wilson chain at the $N$-th step is obtained via,  
	\begin{equation}\label{eqb69}
		H_N = U_N^R \mathcal{E}_N (U_N^L)^\dagger.
	\end{equation}
	Thus, we have to calculate both the left and the right transformation matrices $U^R_N$ and $U^L_N$ at each step, and the iterative Hamiltonian in Eq.\eqref{eqb64} is extended to
	\begin{equation}\label{eqb70}
		H_{N+1}(ks,k's') = \prescript{L}{N+1}{\braket{ks|H_{N+1}|k's'}}_{N+1}^R,
	\end{equation}
	so do the eigenvectors, i.e.,
	\begin{equation}\label{eqb71}
		\begin{aligned}
			\ket{w}_{N+1}^R &= \sum_{ks} U^R(w,ks) \ket{ks}_{N+1}^R, \\
			\prescript{L}{N+1}{\bra{w}} &= \sum_{ks} \prescript{L}{N+1}{\bra{ks}} (U^L)^\dagger (w,ks).
		\end{aligned}
	\end{equation}
	The basis of the left states are defined in a similar way to the right ones. In this way, the transformation in Eq.\eqref{eqb66} is reinterpreted as
	\begin{equation}\label{eqb72}
		A_N^{\{w\}} = (U^R_N)^\dagger A_N^{\{ks\}} U^L_N. 
	\end{equation}
	Meanwhile, the definition of the trace for non-Hermitian cases is naturally extended to an LR-form according to the biorthogonality and completeness relation (Eq.\eqref{eqb58} and Eq.\eqref{eqb61}), i.e.,~\cite{brody2013biorthogonal,Meden_2023}
	\begin{equation}\label{eqb73}
		{\rm Tr}^{\rm LR} A^{\{w\}}_N:=\sum_{w} \prescript{L}{N}{\bra{w}} A^{\{w\}}_N \ket{w}^R_N, 
	\end{equation}
	where the brackets are taken with the left and right states. Then the expectation value of observables is defined in the form as,
	\begin{equation}\label{eqb74}
		\begin{split}
			\braket{A}_N^{\rm LR} &:= \frac{{\rm Tr}^{\rm LR} A_N e^{-\beta H_N}}{Z_N^{\rm LR}} \\
			&= \frac{\sum_{w} \prescript{L}{N}{\bra{w}} A^{\{w\}}_N \ket{w}^R_N e^{-\beta E_{N,w}}}{\sum_{w} e^{-\beta E_{N,w}}}. 
		\end{split}
	\end{equation}
	
	The similar LR-form definition of thermal dynamic quantities for non-Hermitian systems as that in Eq.\eqref{eqb74} has been adopted in a number of recent studies \cite{def}.  We now discuss its applicability for both the $\mathcal{PT}$-symmetric and asymmetric case. 
	
	For the \(\mathcal{PT}\)-symmetric case, there exists a Hermitian and invertible metric operator \(\eta\) that satisfies the pseudo-Hermiticity relation \(H^\dagger = \eta H \eta^{-1}\). This allows for the construction of a new Hilbert space equipped with a modified inner product, \(\braket{\cdot|\cdot}_\eta := \braket{\cdot|\eta|\cdot}\). The trace in this Hilbert space is defined as \({\rm Tr}_\eta A^{\{w\}}_N := {\rm Tr}\ \eta A^{\{w\}}_N \), which is equivalent to our LR-form definition \({\rm Tr}^{\rm LR}\) under pseudo-Hermiticity. 
	
	When $\mathcal{PT}$ symmetry is not spontaneously broken, the energy spectrum will be strictly real. In this case, the metric \(\eta\) is not only invertible but also positive. Thus, the thermodynamic quantities defined in Eq.\eqref{eqb74} are completely valid and is analogous to those in conventional Hermitian systems \cite{moiseyev2011non,brody2013biorthogonal}. Whereas, when the $\mathcal{PT}$ symmetry is spontaneously broken, the eigenenergies will either be real or take place as complex conjugate pairs. Thus, the partition function \(Z_N^{\rm LR}\) still maintains to be real, since we have \(e^{-\beta E} + e^{-\beta E*} = 2e^{-\beta {\rm Re}(E)} {\rm cos[Im}(\beta E)]\). In this case, it should be noted that the Boltzmann weight and the partition function is not positive-definite in general.  Although this is forbidden for closed Hermitian systems, this feature reflects the essence of the non-Hermiticity, which physically originates from open systems with loss or gain. Moreover, for the model and parameters studied in this work, we will show in the following that the partition function remains positive for all energy scales, as indicated by Fig.\ref{fig:thermo-pt}(b).
	
	The third case considers \(\mathcal{PT}\)-asymmetric Hamiltonians. In this case, both \(Z_N^{\rm LR}\) and \(\braket{A}_N^{\rm LR}\) could become complex. For completeness, one can respectively calculate the real and imaginary part of the thermodynamic quantities. Using the NHMCK model as an example, we will show in the following that both the real and imaginary components are meaningful, which provide clear signatures about the YSR state demonstrated above.  Thus, the usefulness of the definition in Eq.\eqref{eqb74} is justified for all three different cases.
	
	Now we discuss more numerical details about solving the left and right eigenstates, which is complicated by degeneracies, as is also discussed comprehensively in Ref.~\cite{19td-1k9s}. The degeneracies often arise from symmetries, either at the Hamiltonian level or the emergent ones at low-energies. The problem can be partially overcome by  decomposing the Hamiltonian into subspaces of different irreducible representations. Then, one only needs to solve certain reduced block matrices. Since the sizes of the blocks are much smaller than the full matrix, the computational cost is also reduced~\cite{weichselbaum_non-abelian_2012,PhysRevB.86.245124}. Taking the $n=2$ NHMCK model as an example, one can utilize the \({\rm U}(1)_{\rm charge}\times {\rm SU}(2)_{\rm channel}\) symmetry.
	However, for symmetries emergent around the low-energy fixed points \cite{affleck1995conformal}, the degeneracy cannot be avoided. For non-Hermitian matrices, most of numerical eigensolver packages do not support a biorthonormal output. One has to orthogonalize and normalize the eigenstates manually in each degenerate subspace. In fact, one can first solve the right eigenvectors with common eigensolvers, and then obtain the left eigenvectors by the inverse of the right one (when invertible), i.e.,
	\begin{gather}\label{eqb7576}
		H\ket{\psi_j}^R = E_j \ket{\psi_j}^R,\\
		U^L = (U^R)^{\dagger-1}.
	\end{gather}
	Utilizing the common eigensolvers as well as the non-Abelian symmetries, the NHMCK can be efficiently solved, as will be discussed below.
	
	Truncation is another key procedure in NRG calculations. In Hermitian cases, the usual scheme is just to drop out excited states higher than an energy cutoff \(E_{\rm k}\), keeping the lowest \(N_{\rm k}\) states.  For non-Hermitian cases, the eigenenergies are generally complex and the cutoff needs a careful benchmark. In Ref.~\cite{19td-1k9s}, several truncation schemes are discussed, and a ``LowRe'' scheme is demonstrated to yield the most stable and accurate numerical results given the cutoff \(N_{\rm k}\). This scheme sorts the eigenvalues primarily by their real parts \(\textrm{Re}(E)\) and secondarily by their imaginary parts \(\textrm{Im}(E)\). 
	We adopt the same cutoff scheme in our calculations in the main text, and provide a basic convergence analysis on different truncation schemes as well as the discretization parameter \(\Lambda\) in Appendix.~\ref{convergence}.
	
	\subsection{Non-Hermitian NRG results for the \textit{PT}-asymmetric model}
	
	We now solve the NHMCK model taking $n=2$ (two-channel) as an example. Compared to models with more channels, it is of minimal computation cost and is already sufficient to demonstrate the key numerical features.  The channel-symmetry is assumed since the proposed setup in Sec.~\ref{sec2} ensures exact symmetry between different channels. For the \(\mathcal{PT}\)-asymmetric model, we consider a general complex Kondo exchange coupling, \(J=J_0e^{i\theta}\) with a tunable \(\theta\) that quantifies the strength of non-Hermiticity. For the \(\mathcal{PT}\)-symmetric model, we assume anisotropic Kondo coupling that takes the form of \({\bf J}=(J_x,J_y,J_z)=J_0(e^{i\theta},e^{-i\theta},1)\). Such an anisotropic model can also be realized by tuning the tunneling coefficient in the multi-junction setup, as discussed earlier.  For \(\theta=0\), both cases reduce to the conventional Hermitian two-channel Kondo model. In this section, we show the NRG results for the \(\mathcal{PT}\)-asymmetric model, while those for the \(\mathcal{PT}\)-symmetric model will be presented in the next section.
	
	\begin{figure*}[htb]
		\centering
		\includegraphics[width=\linewidth]{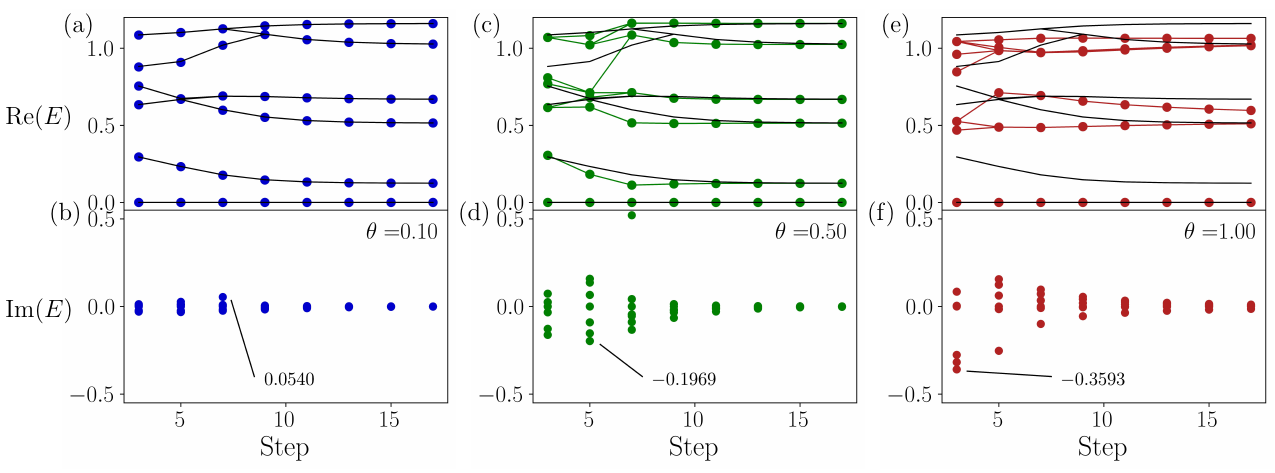}
		\caption{The energy spectrum for the \(\mathcal{PT}\)-broken $n=2$ NHMCK model  obtained by non-Hermitian NRG for different $\theta$. The upper ((a)(c)(e)) and lower ((b)(d)(f)) panel show  the real and imaginary part, i.e., Re\((E)\) and Im\((E)\), respectively. The maximal absolute value of the imaginary parts are marked for each plot in (b)(d)(f). The first 50 eigenenergies are plot here. The black curves show the calculated Re\((E)\) corresponding to the Hermitian two-channel Kondo model with \(\theta=0\).
        For the $\mathcal{PT}$-asymmetric model, we take the numerical parameters as \(\Lambda=6\), \(N_{\rm k}=2048\), and adopt the ``LowRe'' truncation scheme. The same parameters are used in Fig.\ref{fig:thermo-iso}.}
		\label{fig:spec-iso}
	\end{figure*}
	
	In our numerical calculations, we take the Kondo coupling strength \(J_0=0.1\), and the numerical parameters are chosen as, \(\Lambda=6\), \(N_{\rm k}=2048\). We employ a traditional scheme to diagonalize the matrices and apply the Abelian symmetry \({\rm U}(1)_{\rm charge}\). Note that in the matrix product state (MPS) formalism, the \({\rm SU}(2)_{\rm channel}\) non-Abelian symmetry can also be applied to further improve efficiency \cite{weichselbaum_non-abelian_2012,PhysRevB.86.245124}.
	
	The energy spectrum of the \(\mathcal{PT}\)-asymmetric model is shown in Fig.\ref{fig:spec-iso}. 
	Notably, the spectrum deviates from the Hermitian result (the dark curves) as \(\theta\) increases. 
	For small $\theta$ , the real part of the spectrum Re\((E)\) agrees well with that of the Hermitian case (Fig.\ref{fig:spec-iso}(a)), with only a negligible imaginary part (Fig.\ref{fig:spec-iso}(b)). For intermediate $\theta$, Re\((E)\) displays obvious deviations at relatively high energy scales  (Fig.\ref{fig:spec-iso}(c)), exhibiting larger imaginary part in the same energy range (Fig.\ref{fig:spec-iso}(d)). However, the system still flows to the Hermitian fixed point at low energies. For large $\theta$, the spectrum is completely distinct from the Hermitian, and the low-energy fixed point is also altered. The absolute values of the imaginary part Im(\(E\)) also increase with \(\theta\), as marked in Fig.\ref{fig:spec-iso}(b),(d),(f).
	
	\begin{figure*}[htb]
		\centering
		\includegraphics[width=\linewidth]{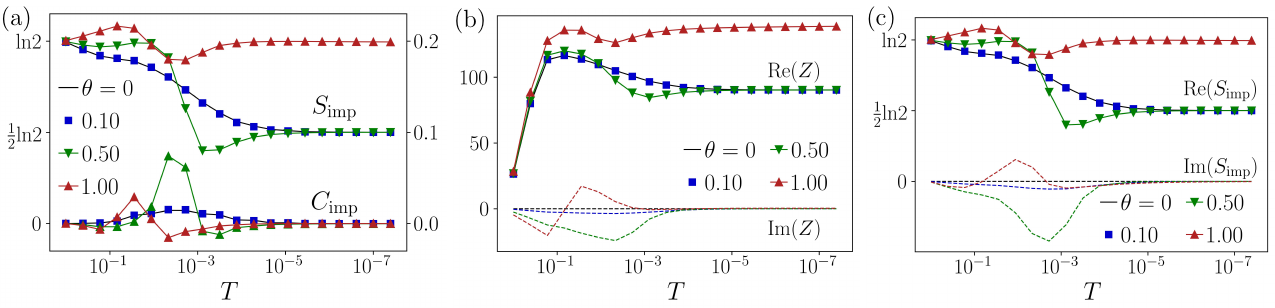}
		\caption{The NHNRG results of thermodynamic quantities as a function of temperature. (a) The calculated impurity entanglement entropy \(S_{\rm imp}\) and impurity specific heat \(C_{\rm imp}\) for the \(\mathcal{PT}\)-asymmetric $n=2$ NHMCK model. (b) The real and imaginary parts of the partition function \(Z_N^{\rm LR}\) and (c) shows those of the impurity entropy \(S_{\rm imp}\). 
		}
		\label{fig:thermo-iso}
	\end{figure*}
	
	Thermodynamic quantities for the \(\mathcal{PT}\)-asymmetric model also display interesting non-Hermitian effects. We calculate the impurity entropy $S_{\rm imp}$ and the impurity specific heat $C_{\rm imp}$ as a function of temperatures. 
    They are defined as the thermodynamic quantities of the total system subtracting that of the bath~\cite{RevModPhys.80.395},
    \begin{align}
    	S_{\rm imp}(T) &= S_{\rm tot}(T) - S_{\rm bath}(T), \\
    	C_{\rm imp}(T) &= C_{\rm tot}(T) - C_{\rm bath}(T),
    \end{align}
    where $S_{\mathrm{tot/bath}}(T)$ and $C_{\mathrm{tot/bath}}(T)$ are computed according to the LR-form defined above.

	As shown in Fig.\ref{fig:thermo-iso}(a), the model in fact exhibits three different regimes as \(\theta\) increases.  The first is the Kondo regime with weak non-Hermiticity (\(\theta=0.10\)) where \(S_{\rm imp}\) monotonically evolves from ln2 to \(\frac{1}{2}\)ln2. This indicates the  (over)screening process of the impurity with lowering temperature, similar to the Hermitian case. 
	The second is the intermediate non-Hermiticity regime (\(\theta=0.50\)), where \(S_{\rm imp}\) also evolves from ln2 to \(\frac{1}{2}\)ln2, but clearly in a non-monotonous behavior. The third is the strong non-Hermiticity regime (\(\theta=1.00\)),  where the impurity entropy  $S_{\rm imp}$ is close to ln 2 in both the UV and IR limit, displaying the local moment phase. 
	
	Remarkably, for both the intermediate and large $\theta$, we observe non-monotonicity, and $S_{\rm imp}$ can overshoot the impurity free moment of ln2 at intermediate temperatures. These behaviors mark a violation of the \(g\)-theorem that the boundary DOF monotonically decrease as the energy scale runs from UV to IR \cite{affleck_exact_1993}.  This is only possible when there emerges extra
	DOF with finite energy, which is a direct signature of the YSR state. 
	
	We notice that similar behaviors were also recently found in the investigation of the impurity thermodynamics at the edge of a single-channel or multi-channel superconducting bath using Bethe ansatz \cite{kattel2025thermodynamicssplithilbertspace,kattel2025multichannelkondoeffectsuperconducting}. 
	In Ref.~\cite{kattel2025thermodynamicssplithilbertspace}, the authors find four phases, including the conventional Kondo phase, the unscreened local moment phase, and the YSR I and YSR II phases. Developed method in Ref.~\cite{kattel2025thermodynamicssplithilbertspace} allows one to extract the full thermodynamics, and it is shown that the impurity entropy overshoots ln2 
	in both YSR phases, which flows to 0 (screened) and ln2 (unscreened) respectively at the IR fixed point. Through analysis of the structure of the towers of eigenstates, the four phases can be identified and separated by boundary quantum phase transitions. The non-monotonicity of \(S_{\rm imp}\) is also discussed in Ref.~\cite{kattel2025thermodynamicssplithilbertspace}, and is attributed to the impurity-induced midgap bound state, namely the YSR mode, whose thermal population causes a transient bump near its energy scale. Our NRG results indicate that similar physics takes place in NHMCK models.
	
	
	We further compare the imaginary parts of the partition function \(Z\) and the impurity entropy \(S_{\rm imp}\) to their real parts in Fig.\ref{fig:thermo-iso}(b) and (c). We find that the imaginary parts vanish as the NRG iteration goes to the IR fixed point, but can not be ignored at higher energies. Furthermore, the nonzero imaginary parts hallmark the characteristic energy scale where real parts of the impurity thermodynamic properties show YSR-like behaviors. The non-monotonic entropy and the overshoot of $\mathrm{ln}2$ indicate that there are extra DOFs with finite energy contributing to the impurity thermodynamics, which are totally induced by the non-Hermitian effects.
	\begin{figure*}[tb]
		\centering
		\includegraphics[width=\linewidth]{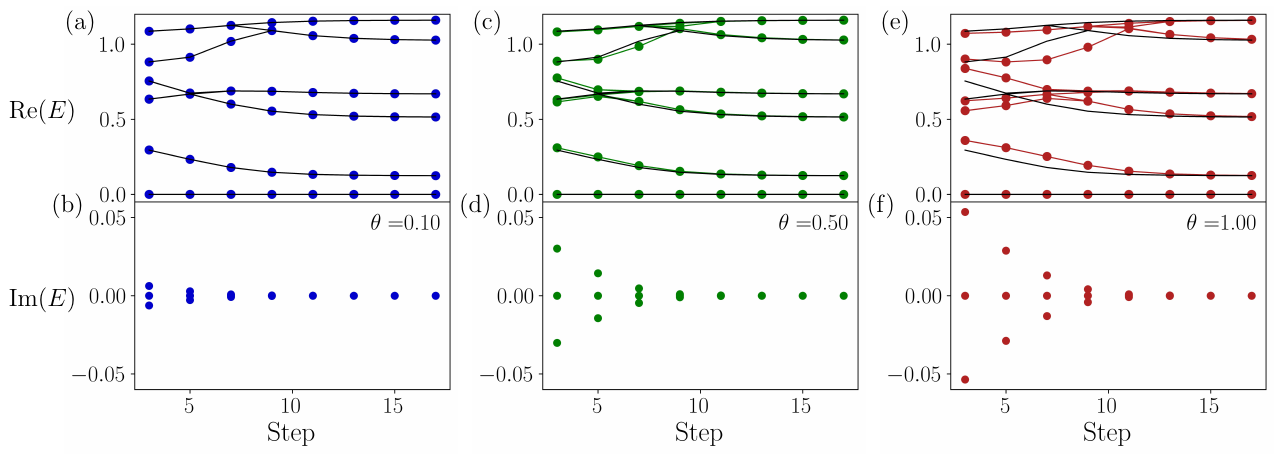}
		\caption{The NHNRG results of the energy spectrum for the \(\mathcal{PT}\)-symmetric $n=2$ NHMCK model for different $\theta$. For the $\mathcal{PT}$-symmetric model, we take the numerical parameters as \(\Lambda=6\), \(N_{\rm k}=2048\), and adopt the ``LowRe'' truncation scheme as well. The same parameters are used in Fig.\ref{fig:thermo-pt}.}
		\label{fig:spec-pt}
	\end{figure*}

	\section{the pseudo-Hermitian model with \textit{PT} symmetry}\label{ptm}
	\subsection{Non-Hermitian NRG results for the \textit{PT}-symmetric model}
	In this section, we discuss the NRG results of the \(\mathcal{PT}\)-symmetric $n=2$ NHMCK model. The \(\mathcal{PT}\) symmetry is ensured by taking the Kondo couplings as, \({\bf J}=J_0(e^{i\theta},e^{-i\theta},1)\). 
	
	The calculated energy spectrum as a function of the RG step is shown in Fig.\ref{fig:spec-pt}. As shown, up to the intermediate $\theta$ (\(\theta=0.50\)), the spectrum almost coincides with that of the Hermitian case with negligible deviations (see Appendix~\ref{thermobcft}). For large $\theta$ (\(\theta=1.00\)), although there occur deviations at higher energy scales, the spectrum exactly flows to that corresponding to the Hermitian fixed point. Thus, for our calculated \(\theta\) up to $\theta=1$, the real part of the spectrum always saturates to the Hermitian case at low temperatures. Meanwhile, at low-energies, the ground state energy remains to be real, with Im\((E_0)<10^{-8}\). As shown, Im\((E_0)\) is at least one order of magnitude smaller than the real part for all energy scales, in sharp contrast with the $\mathcal{PT}$-asymmetric model in Fig.\ref{fig:spec-iso}. Moreover, we find that both the low-energy spectrum and the degeneracies coincide with the predictions of BCFT (see Appendix~\ref{thermobcft}), as shown in Fig.\ref{fig:thermo-pt}(a). Thus, it can be concluded that, with imposing $\mathcal{PT}$ symmetry, the Hermitian strong-coupling fixed point predicted by the perturbative RG analysis in Sec.~\ref{prg} becomes valid, and the BCFT approach can be applied to further explore physical properties associated with the fixed point \cite{zhang_majorana_1999,PhysRevB.45.7918}.
	
	\begin{figure*}[htb]
		\centering
		\includegraphics[width=\linewidth]{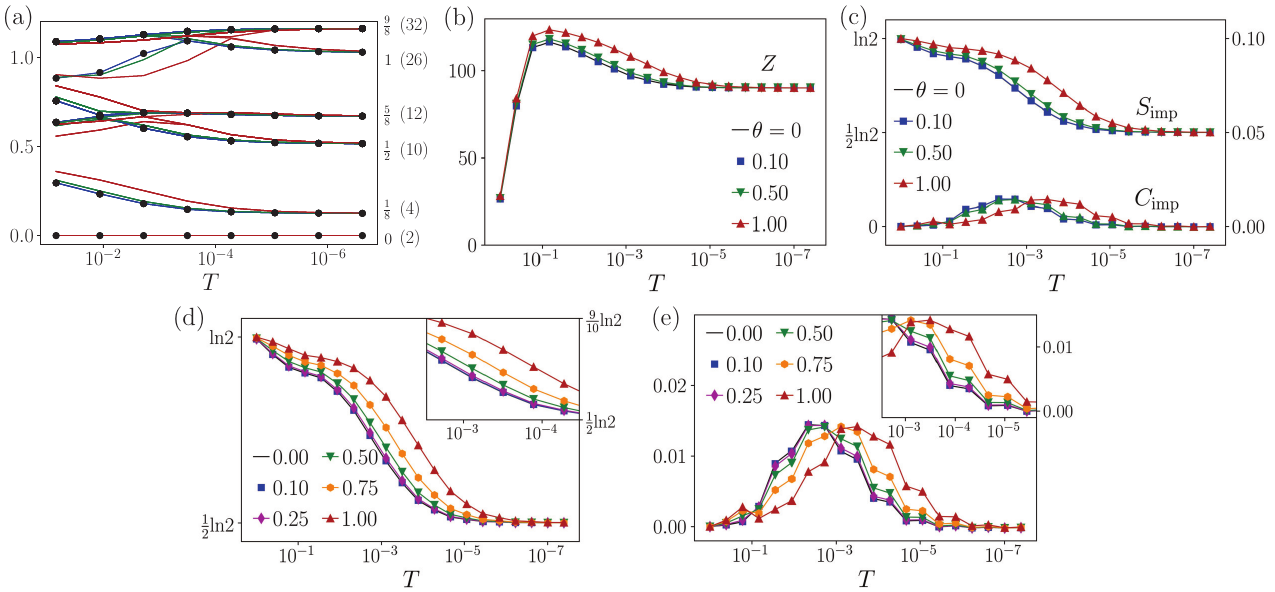}
		\caption{(a) The NHNRG results of the thermodynamic quantities for \(\mathcal{PT}\)-symmetric $n=2$ NHMCK. (a) shows the real part of the energy spectrum, where the dark dot, blue, green, and red curves are the results for \(\theta=0,\ 0.1,\ 0.5,\ 1.0\), respectively. On the right side, we show the energy levels and the degeneracies (in the bracket) predicted by BCFT around the low-energy fixed point, which is in exact agreement with the NRG results. (b) shows the partition function as a function of temperature for various $\theta$, which is positive for all energy scales. (c) The calculated impurity entanglement entropy \(S_{\rm imp}\) and impurity specific heat \(C_{\rm imp}\) as a function of temperature. (d) and (e) are the zoom-in plot of (c) in the intermediate temperature range, which displays non-negligible shifts of thermodynamic quantities due to the non-Hermitian effects.
		}
		\label{fig:thermo-pt}
	\end{figure*}
	
	In Fig.\ref{fig:thermo-pt}(b), we further observe that the partition function always maintains positive for the \(\mathcal{PT}\)-symmetric model, even if the \(\mathcal{PT}\) symmetry can be spontaneously broken.
	In Fig.\ref{fig:thermo-pt}(c), we plot the thermodynamic properties associated with the impurity. As shown, both the impurity entropy \(S_{\rm imp}\) and the impurity specific heat \(C_{\rm imp}\) flow to the strong-coupling overscreening fixed point at low temperatures for various $\theta$. Fig.\ref{fig:thermo-pt}(d) and (e) show the zoom-in data of \(S_{\rm imp}\) and \(C_{\rm imp}\)  with taking more different values of \(\theta\), respectively. Interestingly, we find that the thermodynamic quantities display non-negligible shifts as $\theta$ is enlarged. This indicates that although the non-Hermiticity does not alter the fixed point at low-temperatures, it has nontrivial effects and modify thermodynamic properties at finite temperatures. This will be discussed again in our BCFT calculations in Sec.~\ref{secbcft} and Appendix~\ref{thermobcft}.

	\subsection{Applicability of BCFT under \textit{PT} symmetry}
	The NHNRG method above has shown that the low-energy spectrum of the $\mathcal{PT}$-symmetric model exactly coincides with the BCFT predictions, i.e., in agreement with the spectrum corresponding to the  strongly correlated fixed point.  In this section, we will give a more rigorous proof that  the low-energy theory of
	\(\mathcal{PT}\)-symmetric model is related to the Wess-Zumino-Witten (WZW) CFT via a similarity transformation. Thus, the BCFT method can be readily applied to study the \(\mathcal{PT}\)-symmetric NHMCK models, which can further yield, in a non-perturbative way, thermodynamic and transport properties associated with the strong-coupling fixed point, as will be discussed in Sec.~\ref{trans} below. 
	
	Let us emphasize again, for \(\mathcal{PT}\) symmetric systems involving no exceptional point, the non-Hermitian
	Hamiltonian $\widetilde{H}$ (we slightly shift our notation here for convenience) can be transformed into a Hermitian Hamiltonian
	via a similarity transformation,
	\begin{equation}\label{eqb77}
		H=S\widetilde{H}S^{-1},
	\end{equation}
	where $S$ is not unitary otherwise it reduces to Hermitian case. Note Eq.\eqref{eqb77} can be derived from the pseudo-Hermitian relation with $S\sim\eta^{1/2}$ ~\cite{bender2007making} and the following symmetry condition is satisfied, i.e.,
	\begin{equation}\label{eqb78}
		[\mathcal{PT},\widetilde{H}]=0,
	\end{equation}
	and 
	\begin{equation}\label{eqb79}
		[\mathcal{PT},S]=0,[\mathcal{PT},S^{-1}]=0.
	\end{equation}
	Note that $\widetilde{H}$ and $H$ share the same spectrum, $\{\varepsilon_{n}\}$. Given the corresponding eigenstates $\{\left|\varepsilon_{n}\right\rangle \}$ for $H$, $\{S^{-1}\left|\varepsilon_{n}\right\rangle \}$ are the right eigenstates for $\widetilde{H}$, and the left eigenstates are $\{(S)^{\dagger}\left|\varepsilon_{n}\right\rangle \}$ (which are also the right eigenstates of $\widetilde{H}^{\dagger}$) .
	From another point of view, the \(\mathcal{PT}\) symmetric system is equivalent to a Hermitian
	system with a metric $g_{\mu\nu}=S^{\dagger}S$ inserted in Hilbert space. 
	
	In the \(\mathcal{PT}\) symmetric non-Hermitian setting, the primaries now become $\{S^{-1}\phi_{j}S,S^{-1}\left|\phi_{j}\right\rangle \}$, thus we define 
	\begin{equation}\label{eqb80}
		\widetilde{\phi}_{j}\equiv S^{-1}\phi_{j}S,
	\end{equation}
	and
	\begin{equation}\label{eqb81}
		\ket{\widetilde{\phi}_{j}} \equiv\widetilde{\phi}_{j} \ket{\widetilde{\textrm{vac}}} =(S^{-1}\phi_{j}S)S^{-1} \ket{\textrm{vac}} =S^{-1} \ket{\phi_{j}}. 
	\end{equation}
	Correspondingly, the CFT Hamiltonian, $H\sim L_{0}$, becomes $\widetilde{L}_{0}\equiv S^{-1}L_{0}S$. The  scaling dimension $\Delta_{j}$ remains unchanged because,
	\begin{equation}\label{eqb82}
		\widetilde{L}_{0} \ket{\widetilde{\phi}_{j}} =S^{-1}L_{0}SS^{-1} \ket{\phi_{j}} =\Delta_{j}S^{-1} \ket{\phi_{j}} =\Delta_{j} \ket{\widetilde{\phi}_{j}}.
	\end{equation}
	Using Eq.\eqref{eqb57}, the operator product expansion (OPE) satisfies,  
	\begin{equation}\label{eqb83}
		\begin{aligned}
			{}^{L}\braket{\widetilde{\phi}_{i}\times\widetilde{\phi}_{j}}{}^{R}&= \braket{SS^{-1}\phi_{i}S\times S^{-1}\phi_{j}SS^{-1}} \\ &= \braket{\phi_{i}\times\phi_{j}} =\left\langle\sum_{k}N_{ij}^{k}\phi_{k}\right\rangle.
		\end{aligned}
	\end{equation}
	Defining ${}^{L}\braket{\widetilde{\phi}_{i}\times\widetilde{\phi}_{j}}{}^{R} \equiv {}^{L}\braket{\sum_{k}\widetilde{N}_{ij}^{k}\widetilde{\phi}_{k}}{}^{R}$, we therefore have $N_{ij}^{k}=\widetilde{N}_{ij}^{k}$, i.e., the fusion coefficient remains unchanged.
	
	In a BCFT, similarly, the boundary state $\ket{\widetilde{b}} $ satisfies,
	\begin{equation}\label{eqb84}
		\left(\widetilde{L}_{n}-\widetilde{\overline{L}}_{-n}\right)\ket{\widetilde{b}} =0
	\end{equation}
	where $\widetilde{L}_{n},\widetilde{\overline{L}}_{-n}$ are the generalized Virasoro generators, $\widetilde{L}_{n}=SL_{n}S^{-1},\widetilde{\overline{L}}_{-n}=S\overline{L}_{-n}S^{-1}$. The Ishibashi kets $\|\widetilde{i}\rangle$, i.e., solutions from Eq.\eqref{eqb84} can be constructed from the Hermitian $\|i\rangle$, since $\ket{\widetilde{b}} =S^{-1}\left|b\right\rangle $. Let $\|\widetilde{i}\rangle=S^{-1}\|i\rangle$, then the Cardy states constructed from the $\{\|\widetilde{i}\rangle\}$ basis, $\|\widetilde{B}\rangle=S^{-1}\|B\rangle$, clearly satisfy Eq.\eqref{eqb84}. 
	
	In particular, the overlap
	\begin{equation}\label{eqb85}
		Z_{BB}(q)=\braket{\bra{\widetilde{B}}\!| q^{\frac{1}{2}\left(\widetilde{L}_{0}+\widetilde{\bar{L}}_{0}-\frac{c}{12}\right)} || \widetilde{B}}
		= \sum_{i}\tilde{N}_{i}\chi_{i}(q)
	\end{equation}
	remains unchanged compared to the Hermitian case. Likewise, the Affleck-Ludwig boundary entropy,
	\begin{equation}\label{eqb86}
		g=\langle\langle\widetilde{0}\|\widetilde{B}\rangle=\langle\langle0\|B\rangle,
	\end{equation}
	also coincides with its Hermitian counterpart. Therefore, we have shown that for \(\mathcal{PT}\)-symmetric models without exceptional point, the BCFT approach applies directly.  In addition, when one calculates thermodynamical quantities in the \(\mathcal{PT}\) region, the Boltzmann weight reduces to that of Hermitian case with only a metric inserted in Hilbert space.  Therefore, we have shown that if there is no spontaneous \(\mathcal{PT}\) symmetry breaking, the BCFT description remains soundly applicable in the low-energy window. Also, this is verified by the consistency with the NHNRG results, where the imaginary energy apparently vanishes in the low-energy spectrum, i.e., Fig.\ref{fig:spec-pt}(a). In addition, the thermodynamic quantities such as specific heat and conductance will be discussed in the following sections and Appendix~\ref{thermobcft}.
	
	\section{transport properties around fixed points}\label{trans}
	In this section, we aim to study the transport properties associated with the fixed points of the NHMCK model.  For the weak-coupling fixed point discussed in  Sec.~\ref{prg}, since it is well within the description of perturbation theory, we will calculate the corresponding Kondo conductance based on the non-Hermitian Kubo formula. While for the strong-coupling fixed point, which is essentially non-perturbative, we restrict ourselves to the \(\mathcal{PT}\)-symmetric model and adopt the BCFT approach, whose validity has been justified for the \(\mathcal{PT}\) symmetric case in the previous section.
	
	\subsection{non-Hermitian Kubo formula approach to the weak-coupling fixed point }
	
	We start from the microscopic junction co-tunneling Hamiltonian (Eq.\eqref{eqb17}) as follows,
	\begin{equation}\label{eqb87}
		H= \sum_{\alpha=1}^M \sum_p \epsilon_p \psi_{p,  \alpha}^{\dagger} \psi_{p,  \alpha}+\frac{1}{2} \sum_{\alpha \neq \beta=1}^M \lambda_{\alpha \beta} \gamma_\alpha \gamma_\beta \psi_{\beta}^{\dagger} \psi_{ \alpha},
	\end{equation}
	where  $\alpha,\beta$ are the junction indices. The coupling coefficients $\lambda_{\alpha \beta} $ are complex numbers. Around the weak-coupling fixed point, we take $\lambda_{\alpha \beta}=\lambda_r-i \lambda_i $,  with $|\lambda|\ll1$. Due to the non-Hermiticity of the Hamiltonian, the dynamics is different from the Hermitian case. The state (density matrix) evolution is given by \cite{brody_mixed-state_2012},
	\begin{equation}
		\frac{d \rho}{d t} = -i [H_{r},\rho] - \left( \{ H_i,\rho\} - 2 \text{Tr}
		(\rho H_i) \rho \right), 
		\label{eqb88}
	\end{equation}
	where $H_{r}$ and $H_i$  represent the Hermitian and  anti-Hermitian parts of the original Hamiltonian $H$, respectively, i.e., $H=H_{r}-iH_i$. Given an initial state $\rho_0$, the solution of Eq.\eqref{eqb88} can be expressed in the form as Eq, \eqref{eqb9}, 
	\begin{eqnarray}
		\rho_t = \frac{e^{-{ i}(H_{r}-{ i}{H_i})t} \rho_0 
			e^{{ i}(H_{r}+{ iH_i})t}} 
		{\text{Tr} \left({e^{-{i}(H_{r}-{ i}{H_i})t} \rho_0 
				e^{{ i}(H_{r}+{ i}{H_i})t}}\right)}, 
		\label{eqb89} 
	\end{eqnarray} 
	which provides a natural generalization of its unitary counterpart. Note that with the denominator, the trace of Eq.\eqref{eqb89} is preserved, while the trace of the unnormalized density matrix, $\tilde{\rho}_t=e^{-{ i}(H_{r}-{ i}{H_i})t} \rho_0 
	e^{{ i}(H_{r}+{ i}{\mathit\Gamma})t}$, which is directly derived from the Lindblad master equation, is generally not 1, unless the Hamiltonian always holds a real energy spectrum.  This has been discussed above in Sec.~\ref{sec2}.
	
	Then, the dynamics of  a general operator $A$ in Heisenberg picture is given by, 
	\begin{equation}\label{eqb90}
		\frac{d A}{d t} = i [H_{r},A] - \{{H_i},A\}.
	\end{equation}
	In comparison,  the dynamics of the observable, $\langle A\rangle= \text{Tr}(A \rho_t)$, is however obtained as,  
	\begin{eqnarray}\label{eqb91}
		\frac{d \langle A\rangle}{d t} = i \langle[H_{r},A]\rangle - \langle\{{H_i},A\}
		\rangle + 2\langle{H_i}\rangle \langle A\rangle. 
	\end{eqnarray}
	We focus on the current operator through the $\rho$-th lead, derived as,
	\begin{equation}\label{eqb92}
		\begin{aligned}
			\tilde{I}_{\rho}&=-\frac{d N_{ \rho}}{d t}=-i\left[H_{r},N_{ \rho} \right]-\{{H_i},N_{ \rho} \}\\
			&=-i\left[H_K',\psi_{ \rho}^{\dagger} \psi_{ \rho}\right]+\{H_{K*},\psi_{ \rho}^{\dagger} \psi_{ \rho}\},
		\end{aligned}
	\end{equation}
	where $H_K'$ and $H_{K*}$ represent the real part and the imaginary part of the second-order co-tunneling terms, i.e., the Kondo exchange terms. Note that the operator derived from Eq.\eqref{eqb90} is Hermitian, and its expectation value is real, owing to Eq.\eqref{eqb88} which  guarantees the trace condition and thus the particle conservation. This is consistent with the post-selection on the Lindbladian,  i.e., resetting the particle number, which also ensures the  trace normalization of the density matrix, resulting in the particle number conservation.
	
	After calculating the commutators above, the current operator is cast into the form, 
	\begin{equation}\label{eqb93}
		\tilde{I}_{\rho}=I_{\rho,n}+I_{\rho,a},
	\end{equation}
	where $I_{\rho,n}$ denotes the normal part and $I_{\rho,a}$ is an anomalous part, i.e., 
	\begin{equation}\label{eqb94}
		I_{\rho,n} =  \frac{i}{2}\sum_{\alpha=1}^{M}\lambda_r\gamma_{\alpha}\gamma_{\rho}\left[\psi_{\alpha}^{\dagger}\psi_{\rho}+\psi_{\rho}^{\dagger}\psi_{\alpha}\right],
	\end{equation}
	and
	\begin{equation}\label{eqb95}
		I_{\rho,a}=\frac{1}{2}\sum_{\alpha\beta}\lambda_i\gamma_{\alpha}\gamma_{\rho}\left(\psi_{\rho}^{\dagger}\psi_{\rho}\psi_{\alpha}^{\dagger}\psi_{\beta}+\psi_{\alpha}^{\dagger}\psi_{\beta}\psi_{\rho}^{\dagger}\psi_{\rho}\right),
	\end{equation}
	where the 4-fermion operators emerge at the zeroth order, implying that the anomalous part of the current comes from the correlation effect due to the monitoring and resetting. Following the non-Hermitian Kubo formula in Ref.~[\onlinecite{sticlet_kubo_2022}], we then obtain the conductance as, 
	\begin{equation}\label{eqb96}
		G_{\alpha \beta}=\operatorname{Re} \lim _{\omega \rightarrow 0} \frac{-1}{\hbar \omega} \int_0^{\infty} \mathrm{d} t \mathrm{e}^{\mathrm{i} \omega_{+} t}\left\langle\left[\tilde{I}_\alpha(t), \tilde{I}_\beta(0)\right]\right\rangle_{\sim}
	\end{equation}
	where 
	\begin{widetext}
		\begin{equation}\label{eqb97}
			\left\langle\left[\tilde{I}_\alpha(t), \tilde{I}_\beta(0)\right]\right\rangle_{\sim}\approx
			\operatorname{Tr}\left\{\left[[\tilde{I}_\alpha(t), \tilde{I}_\beta(0)]_{\sim}-\langle \tilde{I}_\alpha(t)\rangle_0\left[e^{i H^{\dagger} t } e^{-i H t}, \tilde{I}_\beta(0)\right]\right] \cdot \rho(0)\right\}/\mathrm{Tr}\rho(0).
		\end{equation}
	\end{widetext}
	Hereby, we have assumed the quasi-unitary state evolution around the fixed point, i.e.,  $\rho(t)\sim\rho(0)/\mathrm{Tr}(\rho(0))$, which is justified because $ \lambda_i$ is small around the weak-coupling fixed point. 
	
	Then, after introducing the notation $\langle\bullet\rangle_0=\operatorname{Tr}\left\{\bullet\rho(0)\right\}/\mathrm{Tr}(\rho(0))$ and expanding the exponential terms,  
	\begin{equation}\label{eqb98}
		\begin{split}
			e^{i H^{\dagger} t } e^{-i H t}&=1+it\left(H^{\dagger}-H\right)\\
			&+\frac{1}{2}\left(2H^{\dagger}H-H^{\dagger{}^{2}}-H^{2}\right)t^2+\ldots,
		\end{split}
	\end{equation}
	we formally obtain, 
	\begin{equation}\label{eqb99}
		\begin{split}
			&\left\langle\left[\tilde{I}_\alpha(t), \tilde{I}_\beta(0)\right]\right\rangle_{\sim} =  \left\langle [ I_{\alpha,r}(t)+I_{\alpha,i}(t),I_{\beta,r}(0)+I_{\beta,i}(0)]\right\rangle_0\\
			&-\braket{\tilde{I}}_0 \braket{\left[1-H_{K*}t+...,I_{\beta,r}(0)+I_{\beta,i}(0)\right]}_0.
		\end{split}
	\end{equation}
	Note that the cross terms like $\tilde{I}_\alpha(t)I_\beta^*(0) $, $H_{K*}I_\beta(0)$ do not have actual contributions to the conductance after the integral in Eq.\eqref{eqb96}, thus can be ignored. Then, we arrive at,
	\begin{equation}\label{eqb100}
		\begin{split}
			&\left\langle\left[\tilde{I}_\alpha(t), \tilde{I}_\beta(0)\right]\right\rangle_{\sim} = \braket{\tilde{I}}_0 \left\langle
			\left[H_{K*}t+...,I_{\beta,i}(0)\right] \right\rangle_0\\
			&+\left\langle [ I_{\alpha,r}(t),I_{\beta,r}(0)]\right\rangle_0+\left\langle [ I_{\alpha,i}(t),I_{\beta,i}(0)]\right\rangle_0.
		\end{split}
	\end{equation}
	Thus, from Eq.\eqref{eqb100}, one knows that the conductance has three terms,
	\begin{equation}\label{eqb101}
		G=G_0+G_*+\delta G_{\langle\rangle},
	\end{equation}
	where $G_0$ is the conventional conductance corresponding to the Hermitian case,  $G_*$ denotes the contribution of imaginary coupling, and  $\delta G_{\langle\rangle}$ comes from the norm correction. Further inserting the specific form of $\rho_r$, $\rho_i$, and $H_{K,*}$, one can readily find that all these three terms are proportional to the square of the exchange coupling around the weak-coupling FP. Specifically, $G_0\sim\lambda^2_r\sim g^2_r$, and $G_*$, $\delta G_{\langle\rangle}\sim\lambda^2_i\sim g^2_i$.  Further utilizing the RG scaling behavior of $g_r$ and $g_i$  around the weak-coupling fixed point, i.e., $g_r,g_i\sim\frac{1}{\ln(T/T_{\mathrm{Kwc}})}$, it is derived that,
	\begin{equation}
		G_{\alpha\beta}(T)\sim \frac{1}{\ln^2(T/T_\mathrm{Kwc})},
	\end{equation}
	where $T_\mathrm{Kwc}$ is an energy scale emergent from the RG equations that is a non-Hermitian counterpart of the  Kondo temperature in the Hermitian case, as defined in Appendix~\ref{ktrg}. 
	
	Interestingly, this Kondo conductance behavior is distinct from previously reported Kondo systems. Although a similar temperature dependence was found for underscreened Kondo models~\cite{posazhennikova_anomalous_2005,coleman_singular_2003,mehta_regular_2005}, the conductance therein saturates to finite values for low temperatures. In addition, although zero conductance occurs at zero temperature in the unscreened channel of anisotropic multichannel Kondo models, Fermi-liquid-type behavior,  i.e., $(T/T_K)^2$, emerges at low temperatures in that case. Thus, the conductance here is quite unique, and it is attributed to a completely different mechanism–the non-Hermiticity induced decoupling of the local moment. The detailed results and comparison of the conductance are shown in Fig.4 of our joint paper~\cite{Letter}.
	
	\subsection{BCFT approach to the strong-coupling fixed point }\label{secbcft}
	We now study the Kondo conductance for the strong-coupling fixed point using the BCFT approach. Although the perturbative methods fail for this fixed point, we have shown that, when imposing the  \(\mathcal{PT}\) symmetry,  the pseudo-Hermiticity allows us to solve it in a similar manner as the Hermitian stereotype.  This leads to the fact that the BCFT results are applicable in this regime, as a deformation of the Hermitian approach. 
	
	To begin with, we transform the low-energy Hamiltonian in Eq.\eqref{eqb25} into the following form, \cite{affleck_exact_1993},
	\begin{equation}\label{eqb103}
		\begin{split}
			H&=\frac{v_{F}}{2\pi}i\int dr\left(\psi_{Lk}^{\dagger}\partial_{r}\psi_{Lk}-\psi_{Rk}^{\dagger }\partial_{r}\psi_{Rk}\right)\\
			&+g\psi_{Lk}^{\dagger }(0)\frac{\vec{\tau}}{2}\psi_{Lk}(0)\cdot\vec{S},
		\end{split}
	\end{equation}
	where the bath fermions are converted into the left- and right-movers, $\psi_{Lk}$ and $\psi_{Rk}$,  $k$ denotes the channel,  and the repeated indices are summed. 
	
	To proceed, we represent Eq.\eqref{eqb103} by the left movers only, and then bosonize the resultant Hamiltonian into a ${\rm U}(1)\times {\rm SU}(2)_n \times {\rm SU}(n)_2$  WZW non-linear sigma-model (NL$\sigma$M). The key to this approach lies in the fact that both terms of Eq.\eqref{eqb103} can be expressed by current density operators involving the charge, spin, and flavor sectors.  In particular, the second term of Eq.\eqref{eqb103}, i.e., the Kondo interaction, can be written as the spin current operator interacting with a local impurity spin at the origin.
	
	The bosonized bath Hamiltonian can be diagonalized at an intermediate value of the coupling parameter \cite{witten_non-abelian_1984,gogolin_bosonization_2004}, indicating the emergence of an infrared stable fixed point described by the ${\rm U}(1)\times {\rm SU}(2)_n \times {\rm SU}(n)_2$ CFT. Further taking into account the impurity, the left- and right-movers are connected by conformal invariant boundary conditions (CIBCs). For example, open boundary conditions take place for the decoupled case, i.e., $\psi_{Lj}(0)=\psi_{Rj}(0)$, while $\psi_{L}(0)=-\psi_{R}(0)$ applies for the strong-coupling case (for the single-channel Kondo model). Considering the boundary conditions, one then arrives at the BCFT description of the low-energy physics in the vicinity of fixed points. Importantly, the CIBCs can be represented by the boundary states \cite{cardy_boundary_1989}, which are in one-to-one correspondence with the conformal towers, i.e., with the primary operators, and thus contain all low-energy information of a BCFT. Then, by further taking into account the effect of the leading irrelevant operators (LIOs), one can obtain the perturbed effective Hamiltonian for $T\ll T_K$, i.e.,
	\begin{equation}\label{eqb104}
		H^{\prime}_{\mathrm{eff}}=H_{\textrm{FP}}+\kappa_O O_{\textrm{LI}},
	\end{equation}
	where $O_{\textrm{LI}}$ denotes the leading irrelevant boundary operators defined in the theory with CIBCs characteristic of the fixed point, and $\kappa_O \sim 1/T_K^{\Delta}$. From Eq.\eqref{eqb104}, we can further extract the thermodynamic properties around the FPs. 
	
	For the multichannel Kondo model studied here, the LIOs can only come from the fermions rather than the local impurity spin (because it is screened). Focusing on charge-conserving perturbations, the ${\rm SU}(2)_n$ spin families are uniquely labeled by the spin of the operator of the lowest scaling dimension, i.e., the spin-$j$ primary with $j = 0,1,2$. Furthermore, the allowed perturbations have to satisfy two criteria: (i) they have to respect \(\mathcal{PT}\) invariance, and (ii) they have to be in a family that can be obtained from the ones of the unperturbed, free fermion theory through the double fusion with the spin- 1/2 primary \cite{affleck_kondo_1991}.
	
	As an example, we now consider the case, $M=3$ and $n=4$, that is most relevant to our main text. The conformal towers of the ${\rm SU}(2)_4$ spin sector are labeled by the spin of the ``highest weight state" (i.e. the lowest energy state). There is one conformal tower for each spin $j$ with,
	\begin{equation}\label{eqb105}
		j=0,1/2,1,\ldots, 2.
	\end{equation}
	These primary fields have zero conformal spin and left and right scaling dimensions,
	\begin{equation}\label{eqb106}
		\Delta_j=\frac{j(j+1)}{6}.
	\end{equation}
	The allowed perturbations around the strong-coupling FP come from the families obtained after the ${\rm SU}(2)_4$ double fusion, which gives,
	\begin{equation}\label{eqb107}
		\left(0_j\otimes\frac{1}{2}_j \right)\otimes\frac{1}{2}_j=0_j,1_j,\qquad\left(2_j\otimes\frac{1}{2}_j \right)\otimes\frac{1}{2}_j=2_j,1_j.
	\end{equation}
	Note that the $j = 1$ primary occurs here, which does not exist in the free fermion spectrum. Thus, anything derived from this family will lead to non-Fermi liquid behaviors. 
	
	
	In the conventional Hermitian case, the time-reversal (TR) symmetry plays the role as the \(\mathcal{PT}\) symmetry for the studied \(\mathcal{PT}\) symmetric model. To find the LIO satisfying the above two criteria, we first note that the spin-$j = 0, 2$ primary operators are both TR invariant. However, they are either trivial or marginal. Then, the $j=1$ primary itself breaks TR symmetry. However, its descendant $\mathbf{J}_{-1}\phi_{j=1}$ preserves TR symmetry while being irrelevant. Here $\mathbf{J}_{-1}$ is the Kac-Moody generator (spin-lowering operator) with scaling dimension 1. The operator, $\mathbf{J}_{-1}\phi_{j=1}$, has total spin 0 and scaling dimension $\Delta=1+\frac{1}{3}=\frac{4}{3}$, leading to non-Fermi liquid behavior at low temperatures.
	
	For the \(\mathcal{PT}\) symmetric non-Hermitian Kondo model, however, in the anti-Hermitian sector, a new irrelevant operator that satisfies \(\mathcal{PT}\) symmetry can be found, i.e., $i\mathbf{J}_{-1} \phi_{j=2}$. Although the Kac-Moody descendant $\mathbf{J}_{-1} \phi_{j=2}$ itself explicitly breaks \(\mathcal{PT}\) symmetry, the imaginary coupling parameter restores the \(\mathcal{PT}\)-invariance of the whole perturbed Hamiltonian. The non-Hermitian boundary interaction legitimizes such anti-Hermitian perturbation terms. This operator has total spin 1 and the scaling dimension $\Delta=2$, and it is the leading order in the anti-Hermitian sector. 
	
	Following the approach in \cite{zazunov_transport_2014,oshikawa_junctions_2006,beri_topological_2012}, the zero-temperature conductance for the Hermitian case is determined by the fixed point alone and is given by, 
	\begin{equation}
		G_{\alpha\beta}(T=0)=\frac{2e^2}{h}(\delta_{\alpha\beta}-\frac{1}{M}),
	\end{equation}
	where $M$ denotes the number of leads attached to the junction. The low-temperature corrections are determined by the LIO, i.e.,
	\begin{equation}\label{eqs82}
		G_{\alpha\beta}=\frac{2e^2}{h}(\delta_{\alpha\beta}-\frac{1}{M})\left[1-a_{\alpha\beta}(T/T_K)^{2\Delta_M-2}\right],
	\end{equation} 
	where $\Delta_M$ is the scaling dimension of the LIO. Note in multi-terminal junction structures, there is an additional symmetry constraint, which requires neutrality on the vertex operators ~\cite{beri_topological_2012}. This brings about the vanishing first-order contribution of the temperature dependence in Eq.\eqref{eqs82}. 
	
	For the non-Hermitian model, the LIO in the anti-Hermitian sector has additional contributions.  For the $M=3$ ($n=4$) case, the total linear conductance is obtained as,
	\begin{equation}
		G_{\alpha \neq\beta}(T)=G_0 \left[ 1-a_{\alpha \beta} \left( T^{\frac{2}{3}}+O(T)\right) +b_{\alpha \beta} T^2\right].
	\end{equation}
	Here, we note that although the order of correction from anti-Hermitian LIO  is higher than that in the Hermitian sector, it has an opposite sign (coming from $(i)^2=-1$). In addition, the coefficient $b_{\alpha \beta}$ could be relatively large in a general sense. Particularly, this would take place for certain bare parameters that generate RG flows mainly along the imaginary axis around the FP. Therefore, this anti-Hermitian LIO will generally result in a non-negligible upturn in the conductance. 
	
	The conductance in the strong-coupling regime is shown by Fig.4(c) of Ref.~\cite{Letter}. As shown, in addition to the non-Fermi liquid feature, an anomalous upturn could generally emerge with in creasing $T$. Such a behavior is again absent in all previously studied Kondo systems~\cite{posazhennikova_anomalous_2005,coleman_singular_2003,mehta_regular_2005,PhysRevB.83.245308,andrei_fermi-_1995}, serving as a characteristic feature driven by the intertwined effects of non-Hermitian and non-Fermi liquid physics. 
	
	However, we mention briefly that, for the $n=2$ NHMCK, the non-Hermitian effect in IR limit is much weaker than the $n=4$ case. This is because the possible leading operators in anti-Hermitian sector therein turn out to be one-level higher descendants (sub-sub-leading terms), e.g., $\sim i\mathbf{J}_{-2}\cdot\phi$,  with the scaling dimension $\Delta=5/2$. Nevertheless, this would still lead to non-negligible modifications to the conductance and thermodynamic quantities at finite temperatures. Interestingly, the latter is indeed numerically observed by our non-Hermitian NRG calculations, as shown by Fig.\ref{fig:thermo-pt}. Thus, in this aspect, the BCFT and the NHNRG results are again in consistency with each other.

	\section{Conclusion and Discussion}\label{conc}
	
	We remark that although the Kondo systems are studied in this work, the results here imply that exotic physics could also take place for other non-Hermitian correlated systems without quasi-particle descriptions, for example, the non-Hermitian Luttinger liquid ~\cite{lee_kondo_1992,frojdh_kondo_1995,furusaki_kondo_1994,han_complex_2023,tang_reclaiming_2024}. Furthermore, regarding the experimental realization, we would like to highlight again that cold-atom systems provide an alternative and maybe more feasible platform to realize our predicted Kondo phenomena. Recently, it is reported that the multi-orbital cold atoms can be engineered to host both localized and itinerant degrees of freedom~\cite{Gorshkov_2010,PhysRevLett.120.143601}, where the dissipation can also be readily introduced~\cite{PhysRevLett.124.203201,nakagawa_non-hermitian_2018}. Designable dissipation and system-integrable monitoring equipment are also achievable in this platform ~\cite{doi:10.1126/science.1155309,10.1093/ptep/ptaa094,Bouganne_2020,skin_2025,universal_2025,Kondozeno_2025}. Last, since we restrict ourselves to \(\mathcal{PT}\)-symmetric case when calculating the Kondo conductance around the strong-coupling fixed point, it is important to loosen the \(\mathcal{PT}\) restriction and investigate how the conductance (and also the other thermodynamic quantities) would be modified by advanced methods, such as Ref.~\cite{kattel2025thermodynamicssplithilbertspace}. This points to an interesting direction, i.e., the symmetry-resolved non-Hermitian effects in correlated systems.

	\section*{Acknowledgments}
	W. Z. Y and Y. C contributed equally to this work. We acknowledge Wei-Qiang Chen for inspiring discussions. This work was supported by the National Natural Science Foundation of China (No.12274206, No.12322402, No.125B2076), the Scientific Research Innovation Capability Support Project for Young Faculty (SRICSPYF-ZY2025164), the National R\&D Program of China (2024YFA1410500, 2022YFA1403601),  the Quantum Science and Technology-National Science and Technology Major Project (Grant No.2021ZD0302800), the Natural Science Foundation of Jiangsu Province (No.BK20233001), and the Fundamental Research Funds for the Central Universities (KG202501).

	\appendix
	\section{effective non-Hermitian Hamiltonian via coupling to fermionic environment}\label{fen}
	The single-particle leakage of the QD into the environment could be realized by its appropriate coupling to the environment. In this Appendix, we show that from the perspective of conventional many-body theory, the environment also renormalizes the QD,  generating a complex self-energy and an effective imaginary chemical potential~\cite{PhysRevB.109.235139}. This provides a supplemental understanding to the Lindblad framework presented in the main text.
	
	Consider a quantum dot coupled to a fermionic bath. The action of the quantum dot is,
	\begin{equation}
		S_{\mathrm{QD}}=\int dt d^{\dagger}(i\partial_t-\varepsilon_d)d,
	\end{equation}
	with $\varepsilon_d$ being the QD energy level, and the terms involving the environment fermions ($f$, $f^{\dagger}_k$) are described by, 
	\begin{equation}
		S_{\mathrm{coupling}}=V\int dt\sum_k[(f^{\dagger}_kd+d^{\dagger}f_{k})],
	\end{equation}
	where $\epsilon_k$ denotes the bath continuum, and $V$ is the coupling strength. The total action of the QD-environment system is, 
	\begin{equation}
		\begin{split}
			S&=\int dt d^{\dagger}(i\partial_t-\varepsilon_d)d\\
			&+V\int dt\sum_k[(f^{\dagger}_kd+d^{\dagger}f_{k})+f^{\dagger}_k(i\partial_t-\epsilon_k)f_k]
		\end{split}
	\end{equation}
	Since all terms are bilinear, the bath fermions can be integrated out, leading to the effective renormalized QD action, i.e., 
	\begin{equation}
		S^{\mathrm{eff}}_{\mathrm{QD}}=\int dt d^{\dagger}(i\partial_t-\varepsilon_d+|V|^2\sum_k\frac{1}{i\partial_t-\epsilon_k})d.
	\end{equation}
	Defining the shifted QD energy, $\tilde{\varepsilon}_d=\varepsilon_d-\mathrm{Re}\Sigma(0)$, where $\Sigma(0)$ is the self-energy at the Fermi energy $E_F=0$, 
	\begin{equation}
		\Sigma(\omega)=|V|^2\sum_k\frac{1}{\omega-\epsilon_k}
	\end{equation}
	and the broadening function $\Gamma/2=-\pi|V|^2\rho(0)$ with $\rho(0)$ being the density of states at $E_F$, the renormalized QD Green's function is obtained as, 
	\begin{equation}
		G^{\mathrm{eff}}_{\mathrm{QD}}=\frac{1}{i\partial_t-\tilde{\varepsilon}_d+i\frac{\Gamma}{2}}.
	\end{equation}
	From renormalized QD Green's function, it is known that an imaginary chemical potential term is generated.

	\section{non-Hermitian Majorana tunneling via dissipative quantum dot}\label{tunneling}
	The junction Hamiltonian describing the dissipative QD-assisted Majorana-lead tunneling is given by,
	\begin{equation}
		H_{\mathrm{junc}}= H_{\mathrm{QD,eff}}+H_{\mathrm{s-QD}}+ H_{\mathrm{s}}.
	\end{equation}
	As discussed in the main text, $H_{s}$ is the junction Hamiltonian without assistance of the dissipative QD, i.e., $H_{\mathrm{s}}=(t\gamma\psi^{\dagger}+h.c.)+H_{\mathrm{lead}}$, and $H_{\mathrm{s-QD}}$ describes the terms involving the QD, i.e., $H_{\mathrm{s-QD}}=t_{Q}(\gamma+\psi^{\dagger})d+h.c.$
	
	The Schrodinger equation corresponding to the total junction system is
	\[
	i\frac{d\left|\Psi(t)\right\rangle }{dt}=H_{\textrm{junc}}\left|\Psi(t)\right\rangle.
	\]
	In the direct product basis of $\left|n_{s}\right\rangle $ and $\left|n_{d}\right\rangle $, we obtain
	\begin{equation}
		\begin{aligned}i\frac{d\left\langle n_{s}|\Psi(t)\right\rangle }{dt} & =\mathop{\underset{s'}{\sum}}\mathcal{H}_{ss'}\left\langle n_{s'}|\Psi_{\mathbf{}}(t)\right\rangle +\mathop{\underset{\mathop{d}}{\sum}}\mathcal{H}_{sd}\left\langle n_{d}|\Psi_{\mathbf{}}(t)\right\rangle, \\
			i\frac{d\left\langle n_{\mathbf{d}}|\Psi(t)\right\rangle }{dt} & =\mathop{\underset{d'}{\sum}}\mathcal{H}_{dd'}\left\langle n_{\mathbf{d'}}|\Psi_{\mathbf{}}(t)\right\rangle +\mathop{\underset{\mathop{d}}{\sum}}\mathcal{H}_{ds}\left\langle n_{s}|\Psi(t)\right\rangle.
		\end{aligned}
	\end{equation}
	Thus, the Hamiltonian can be written in matrix form, and the entries $\mathcal{H}_{ss'}$, $\mathcal{H}_{dd'}$
	and $H_{sd}$ describes the system, quantum dot and cross term respectively.
	
	For large systems, assuming that the wavefunction has a uniform phase factor $\exp(-iEt)$, the coupled equations are cast into
	\begin{equation}
		E\left\langle n_{s}|\Psi_{\mathbf{}}(t)\right\rangle =\mathop{\underset{s'}{\sum}}\mathcal{H}_{ss'}\left\langle n_{s'}|\Psi(t)\right\rangle +\mathop{\underset{\mathop{d}}{\sum}}\mathcal{H}_{sd}\left\langle n_{d}|\Psi(t)\right\rangle 
	\end{equation}
	and
	\begin{equation}
		E\left\langle n_{d}|\Psi(t)\right\rangle =\mathop{\underset{d'}{\sum}}\mathcal{H}_{dd'}\left\langle n_{\mathbf{d'}}|\Psi(t)\right\rangle +\mathop{\underset{\mathop{d}}{\sum}}\mathcal{H}_{ds}\left\langle n_{s}|\Psi(t)\right\rangle 
	\end{equation}
	Eliminating $\left\langle n_{d'}|\Psi(t)\right\rangle$, one then obtains
	\begin{equation}
		\begin{split}
			E\left|s\right\rangle& =H_{\textrm{eff}}\left|s\right\rangle =H_{s}\left|s\right\rangle +H_{\textrm{s-QD}}(E\mathbb{I}-H_{\textrm{QD}})^{-1}H_{\textrm{s-QD}}^{T}\left|s\right\rangle\\
			&=(H_{s}+H_{\textrm{s-QD}}(E\mathbb{I}-H_{\textrm{QD}})^{-1}H_{\textrm{s-QD}}^{T})\left|s\right\rangle, 
		\end{split}
	\end{equation}
	where we find the effective junction Hamiltonian (after integrating out the dissipative quantum dot) to be,
	\begin{equation}\label{eqs29}
		H_{\textrm{eff}}=H_{s}+H_{\textrm{s-QD}}(E\mathbb{I}-H_{\textrm{QD}})^{-1}H_{\textrm{s-QD}}^{T}.
	\end{equation}
	Thus, the Majorana tunneling is renormalized, leading to
	\begin{equation}
		t^{\prime}_+=t+\frac{|t_{\mathrm{Q}}|^2}{E-\tilde{\varepsilon}_d+i\Gamma/2},~~~~ t^{\prime}_-=t^{\star}+\frac{|t_{\mathrm{Q}}|^2}{E-\tilde{\varepsilon}_d+i\Gamma/2}.
	\end{equation}
	
	\section{critical line and Kondo temperature from RG}\label{ktrg}
	In this Appendix, we present more details of determining the critical line separating the strong- and weak-coupling fixed point in the perturbative RG theory. Starting from the RG flow in Eq.\eqref{eqb22} and integrating the energy cutoff from $\Lambda_0$ to $\Lambda$, we obtain
	\begin{equation}
		\begin{split}
			\int_{\Lambda_0}^{\Lambda}d\ln \Lambda&=\int_{g(\Lambda_0)}^{g(\Lambda)}\frac{dg}{g^2-\frac{n}{2}g^3}\\
			&=-\frac{1}{g(\Lambda)}+\frac{n}{2}\ln g(\Lambda)-\frac{n}{2}\ln (1-\frac{n}{2}g(\Lambda))\\
			&+\frac{1}{g(\Lambda_0)}-\frac{n}{2}\ln g(\Lambda_0)+\frac{n}{2}\ln (1-\frac{n}{2}g(\Lambda_0)).\\
		\end{split}
	\end{equation}
	On the critical line, the coupling $g$ in the complex plane flows towards $g_r(\Lambda)\rightarrow \frac{2}{3n},\quad g_i(\Lambda)\rightarrow \infty$. Using $\mathrm{Re}(\ln z)=\ln\sqrt{x^2+y^2},\quad \mathrm{Im}(\ln z)=\arctan(\frac{y}{x}),\quad \mathrm{Re}(\frac{1}{z})=\frac{x}{x^2+y^2}, \quad  \mathrm{Im}(\frac{1}{z})=-\frac{y}{x^2+y^2}$ , and noting that the energy shell is real although the total energy could be complex, we arrive at,
	\begin{equation}\label{eqs36}
		\begin{split}
			&0=\frac{g_i(\Lambda)}{|g(\Lambda)|^2}+\frac{n}{2}\arctan(\frac{g_i(\Lambda)}{g_r(\Lambda)})-\frac{n}{2}\arctan(\frac{\frac{-ng_i(\Lambda)}{2}}{1-\frac{ng_r(\Lambda)}{2}})\\
			&- \frac{g_i(\Lambda_0)}{|g(\Lambda_0)|^2}-\frac{n}{2}\arctan(\frac{g_i(\Lambda_0)}{g_r(\Lambda_0)})+\frac{n}{2}\arctan(\frac{\frac{-ng_i(\Lambda_0)}{2}}{1-\frac{ng_r(\Lambda_0)}{2}}),
		\end{split}
	\end{equation}
	which can be further simplified to,
	\begin{equation}
		\begin{split}
			0=&\frac{n\pi}{2}-\frac{g_i(\Lambda_0)}{|g_r(\Lambda_0)|^2}-\frac{n}{2}\arctan(\frac{g_i(\Lambda_0)}{g_r(\Lambda_0)})\\
			&+\frac{n}{2}\arctan(\frac{\frac{-ng_i(\Lambda_0)}{2}}{1-\frac{n}{2}g_r(\Lambda_0)}).
		\end{split}
	\end{equation}
	This equation determines the critical line, as plotted in Fig.3(b) of Ref.~\cite{Letter}. The real part of RG flow equation is similarly derived as,
	\begin{equation}
		\begin{split}
			\ln \frac{\Lambda}{\Lambda_0}&=-\frac{g_r(\Lambda)}{|g(\Lambda)|^2}+\frac{n}{2}\ln|g(\Lambda)|+\frac{g_r(\Lambda_0)}{|g(\Lambda_0)|^2}+\frac{n}{2}\ln|g(\Lambda_0)|\\
			&-\frac{n}{2}\ln|1-\frac{n}{2}g(\Lambda)|-\frac{n}{2}\ln|1-\frac{n}{2}g(\Lambda_0)|.
		\end{split}
	\end{equation}
	Now we discuss the Kondo temperature $T_K$ in the non-Hermitian models. For the flow towards the weak-coupling fixed point, a temperature $T_\mathrm{Kdis}$ is defined by Eq.(5) in Ref.~\cite{nakagawa_non-hermitian_2018}, which characterizes the energy scale where the dissipative effect manifests. Here we define a strengthened energy scale $T_\mathrm{Kwc}$, as $\frac{dg_i}{d\ln T_\mathrm{Kwc}}=0$, where the circling flow starts to flow back towards the weak-coupling fixed point. The insertion of Eq.\eqref{eqs36} generates, 
	\begin{equation}
		\begin{split}
			\ln (\frac{T_\mathrm{Kwc}}{\Lambda_0})&=-\frac{3n}{4}+\frac{n}{2}\ln \frac{4}{n}+\frac{g_r(\Lambda_0)}{|g(\Lambda_0)|^2}+\frac{n}{2}\ln|g(\Lambda_0)|\\
			&-\frac{n}{2}\ln|1-\frac{n}{2}g(\Lambda_0)|.
		\end{split}
	\end{equation}
	Therefore, $T_\mathrm{Kwc}$ is given by,
	\begin{equation}
		\begin{split}
			T_\mathrm{Kwc}&=\Lambda_0e^{-\frac{3n}{4}}(4/n)^{n/2}|g(\Lambda_0)|^{\frac{n}{2}}|1-\frac{1}{2}g(\Lambda_0)|^{\frac{-n}{2}}\\
			&\times\exp(\frac{g_r(\Lambda_0)}{|g(\Lambda_0)|^2}).
		\end{split}
	\end{equation}
	Note that near the critical line and for $|g(\Lambda_0) |\ll 1$, this expression is simplified to
	\begin{equation}
		T_\mathrm{Kwc}\sim \Lambda_0e^{-\frac{3n}{4}}(4/n)^{n/2}|g(\Lambda_0)|^{\frac{n}{2}}\exp(\frac{g_r(\Lambda_0)}{|g(\Lambda_0)|^2}).
	\end{equation}
	For $g_i(\Lambda_0)=0$, the above $T_\mathrm{Kwc}$ is further reduced to $T_\mathrm{Kwc}\sim \Lambda_0\sqrt{g(\Lambda_0)}\exp (\frac{1}{g(\Lambda_0)})$, which is of the standard form of the Kondo temperature in the Hermitian cases.
	
	\section{Convergence analysis of NHNRG method}\label{convergence}
	
	\begin{figure*}[htb]
	\centering
	\includegraphics[width=\linewidth]{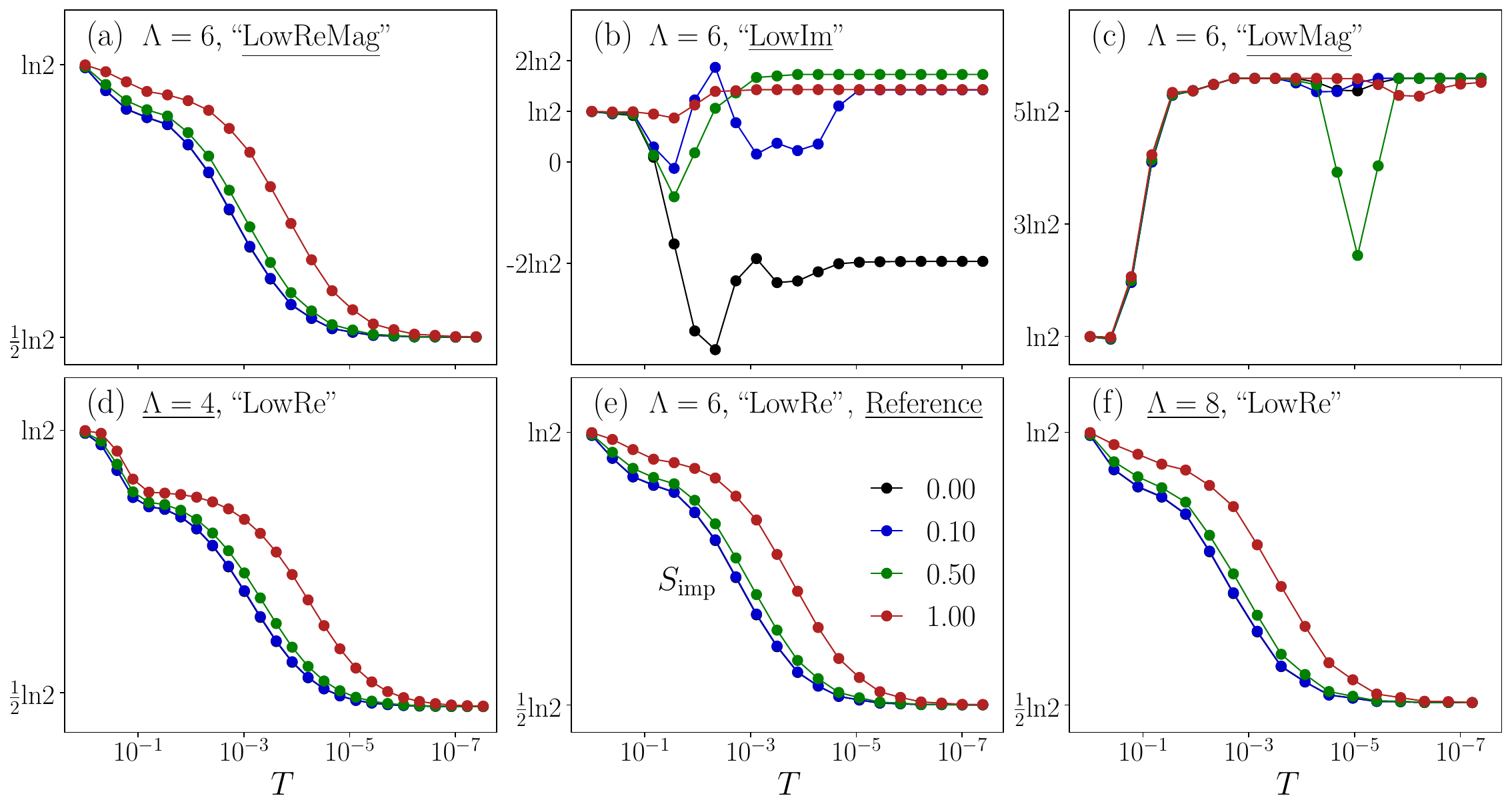}
	\caption{Convergence analysis of the NHNRG method. Impurity entropy $S_{\rm imp}$ is calculated with different values of scaling parameter $\Lambda$ and different truncation schemes. (a-c) are results obtained using the ``LowReMag'', ``LowIm'', and ``LowMag'' schemes respectively, and $\Lambda=6$. (d-f) show the `LowRe'' results for \(\Lambda=4\), 6, 8, respectively. The data in panel (e) are extracted  from Fig.~\ref{fig:thermo-pt}(c) in the main text, and serves as a reference here. The number of kept states is $N_{\rm k}=2048$ for all panels.}
	\label{fig:conv}
\end{figure*}

In this appendix, we perform a convergence analysis of our numerical method. Taking the  $\mathcal{PT}$-symmetric two-channel Kondo model as an example,  we calculate the impurity entropy $S_{\rm imp}$ using different truncation schemes (Fig.\ref{fig:conv}(a)-(c)) and different values of the scaling parameter $\Lambda$ (Fig.\ref{fig:conv}(d)-(f)). Results in Fig.~\ref{fig:conv}(e) are extracted from  Fig.~\ref{fig:thermo-pt}(c), which are obtained with the ``LowRe'' truncation scheme and \(\Lambda=6\) ,  and can be used as a reference to compare with other numerical settings.

In addition to the ``LowRe'' scheme mentioned in the main text, we have testified three other trunction schemes, as were originally introduced in Ref.~\cite{19td-1k9s}.   In the ``LowIm'' scheme, eigenenergies are sorted primarily by their imaginary parts and secondarily by their real parts. In the ``LowMag'' scheme, eigenenergies are sorted simply by their magnitudes. In the hybrid ``LowReMag'' scheme, one firstly searches for the ground state energy within the ``LowRe'' scheme, and then sorts the excited eigenenergies using the ``LowMag'' scheme. 

We compare the results obtained from different truncation schemes for a fixed scaling parameter with \(\Lambda=6\).
As shown, the data obtained with the ``LowReMag'' scheme (Fig.~\ref{fig:conv}(a)) are well converged and agree with those from the ``LowRe'' scheme employed in the main text (Fig.~\ref{fig:thermo-pt}(c) and Fig.~\ref{fig:conv}(e)).  However, under the ``LowIm'' scheme (Fig.~\ref{fig:conv}(b)), the  impurity entropy fails to approach towards the common fixed points for all values of $\theta$. The values of $S_{\rm imp}$ at low temperatures exhibit randomness and cannot even be reduced to the Hermitian result for  $\theta=0.00$. Similarly, under the ``LowMag'' scheme (Fig.~\ref{fig:conv}(c)), the system flows to a new fixed point, where the infrared impurity entropy saturates at $S_{\rm imp}(T\to 0)\approx 5.5{\rm ln}2$, which is much greater than the local moment value. Thus,  the ``LowIm'' and ``LowMag'' scheme either lack  the numerical convergence or fail to produce reliable physical results. 

The above results are in consistence with Ref.~\cite{19td-1k9s}, which remarks that both the ``LowRe'' and ``LowReMag'' schemes perform well and are numerically stable, whereas the ``LowIm'' and ``LowMag'' schemes fail to produce reliable outcomes.

In Fig.~\ref{fig:conv}(d)-(f), we show the impurity entropy obtained using the ``LowRe'' scheme for different scaling parameters, $\Lambda=4,\ 6,\ 8$. It can be found the results are not  affected by the choice of \(\Lambda\), which only displays minor quantitative difference at high temperatures. This aligns with the conventional Hermitian NRG method.

	\section{Thermodynamics analyzed by BCFT}\label{thermobcft}
	In this appendix section, we will give more BCFT details of the thermodynamics of \(\mathcal{PT}\)-symmetric model.
	
	First, the low-energy spectrum of $\mathcal{PT}$-symmetric $n$-channel Kondo model can be given by BCFT as~\cite{zhang_majorana_1999,ludwig_exact_1994},
	\begin{equation}
		E=\frac{\pi}{l}\left[\frac{Q^2}{4n}+\frac{j(j+1)}{n+2}+\frac{j_f\left(j_f+1\right)}{n+2}+n_Q+n_s+n_f\right],
	\end{equation}
	where $Q$ is the total charge of fermions (relative to the ground state), 
	$j$ and $j_f$ are integers or half-integers, and $n_Q, n_s,n_f$ denote the charge, spin, flavor index in conformal tower, respectively, which are non-negative integers. 
	
	For example, when $n=2$, one can obtain several lowest energy levels (in unit $El/\pi$) as follows,
	\begin{equation}
		0,1/8,1/2,1/2+1/8,1,1+1/8,...
	\end{equation}
	This is in exact agreement with that obtained by non-Hermitian NRG in Fig.\ref{fig:thermo-pt}(a).
	
	The leading irrelevant operator in this case, $\mathbf{J}_{-1} \phi_{j}$ contributes to the impurity free energy in the following scaling rule,
	\begin{equation}
		f_{\mathrm{imp}}\propto T^{1+2\Delta},
	\end{equation}
	where $\Delta=\frac{j(j+1)}{n+2}$ denotes the scaling dimension of primary $\phi_j$. Then, the impurity specific heat can be calculated via,
	\begin{equation}
		C_{\mathrm{imp}}=-T\frac{\partial^{2}f_{\mathrm{imp}}}{\partial T^{2}}.
	\end{equation}
	For $n>2$ one obtains,
	\begin{equation}
		C_{\mathrm{imp}}=-T\frac{\partial^{2}f_{\mathrm{imp}}}{\partial T^{2}}\propto\kappa_{1}^{2}\left[T^{2\Delta}+O(T)\right]+O\left(\kappa_{1}^{3}T^{3\Delta}\right)+\ldots
	\end{equation}
	For the case $n=2$, where $\Delta=1/2$, we obtain,
	\begin{equation}
		C_{\mathrm{imp}}\propto\kappa_{1}^{2}T\ln\left(T_{K}/T\right)+O(T)+\ldots
	\end{equation}
	Similarly, the impurity susceptibility scales as $(n>2)$,
	\begin{equation}
		\chi_{\mathrm{imp}}\propto\kappa^{2}T^{2\Delta-1}.
	\end{equation}
	
	Specially, for $n=2$, the possible leading operator in anti-Hermitian sector comes from one-level higher descendant, e.g., $\sim i\mathbf{J}_{-2}\cdot\phi$,  with the scaling dimension $\Delta=5/2$.  This non-Hermitian irrelevant operator leads to corrections of thermodynamic quantities at higher temperatures. In terms of the impurity specific heat, it brings about a correction, 
	\begin{equation}\label{T3}
		C^{\mathrm{nH}}_{\mathrm{imp}}=bT^{3}+...,
	\end{equation}
	on top of the specific heat of the Hermitian model, i.e., 
	\begin{equation}
		C_{\mathrm{imp}}=aT\ln\left(T_{K}/T\right)+O(T)+O(T^{3/2})+...
	\end{equation}
	where $a,b$ are non-universal dimensionless parameters.  Notably, from the NRG calculation in Fig.\ref{fig:thermo-pt}(d)(e), we can see that the non-Hermicity has little effect on the specific heat at low temperatures, which makes sense because we know from above that the non-Hermitian effect is a higher order correction. However, in the relatively high temperature regime,  the numerically extracted specific heat indeed shows increasingly evident deviations from the Hermitian case, this reflects the effect of the non-Hermicity induced   $bT^{3}$ term in Eq.\eqref{T3}. Thus, the NHNRG calculations are consistent with the BCFT predictions.

%

\end{document}